\documentclass[aps,prb,twocolumn,superscriptaddress,nofootinbib]{revtex4-1}
\pdfoutput=1
\usepackage{amssymb,graphicx,color}
\usepackage[intlimits]{amsmath}
\usepackage[english]{babel}
\usepackage[colorlinks]{hyperref}
\usepackage{comment}
\usepackage{tabularx}
\usepackage{array}
\usepackage[svgnames,table]{xcolor}
\usepackage{ragged2e}
\usepackage[normalem]{ulem}

\newcolumntype{A}{>{\centering\arraybackslash \columncolor{white!50!white}}m{2.1cm}}
\newcolumntype{B}{>{\centering\arraybackslash \columncolor{white}}m{7.9cm}}
\newcolumntype{C}{>{\centering\arraybackslash \columncolor{white!50}}m{7.9cm}}
\newcolumntype{D}{>{\centering\arraybackslash \columncolor{white!42}}m}
\newcolumntype{P}[1]{>{\centering\arraybackslash}p{#1}}
\def\beq{\begin{equation}}
\def\eeq{\end{equation}}
\def\bea{\begin{eqnarray}}
\def\eea{\end{eqnarray}}
\def\barr{\begin{array}}
\def\earr{\end{array}}

\begin{document}

\title{Drive-induced many-body localization and coherent destruction of Stark many-body localization  }

\author{Devendra Singh Bhakuni}
\affiliation{Indian Institute of Science Education and Research Bhopal 462066 India}
\author{Ritu Nehra}
\affiliation{Indian Institute of Science Education and Research Bhopal 462066 India}
\author{Auditya Sharma}
%\email{auditya@iiserb.ac.in}
\affiliation{Indian Institute of Science Education and Research Bhopal 462066 India}

\begin{abstract}
%Subjecting a generic clean many-body system to a high-frequency drive leads to a 'infinite temperature-like' state. Here, we show that the application of an electric field drive to a clean many-body syatem can offer a rich variety of phemomena and avoid the featureless infinite temperature-like state.    
We study the phenomenon of many-body localization (MBL) in an
interacting system subjected to a combined dc as well as square wave
ac electric field. % First, the condition for the dynamic localization, coherent destruction of Wannier-Stark localization and super Bloch oscillations in the non-interacting limit, are obtained semi-classically.
First, the condition for the dynamic localization and coherent destruction of Wannier-Stark localization in the non-interacting limit, are obtained semi-classically. In the presence
of interactions (and a confining/disordered potential), a static field alone leads to ``Stark-MBL", for sufficiently large field strengths. We find that in the presence of an additional
high-frequency ac field, there are two ways of maintaining the MBL
intact: either by resonant drive where the ratio of amplitude to the
frequency of the drive ($A/\omega$) is tuned at the dynamic
localization point of the non-interacting limit, or by off-resonant
drive. Remarkably, resonant drive with $A/\omega$ tuned away from the
dynamic localization point leads to a \emph{coherent destruction of
  Stark-MBL}. Moreover, a pure (high-frequency) ac field can also give
rise to the MBL phase if $A/\omega$ is tuned at the dynamic
localization point of the zero dc field problem.
\end{abstract}

\maketitle

\begin{comment}
Driven systems exhibit many fascinating phenomena
ranging from parametric resonance**Ref***, to dynamical stabilization ***Ref** and dynamical
localization~\cite{dunlap1986dynamic,dunlap1988dynamic}. Striking
examples of such systems include the Kapitza pendulum~\citep{kapitza1965dynamical},
kicked rotors~\citep{casati1979stochastic,fishman1982chaos}, and Floquet
topological
insulators~\cite{PhysRevLett.110.200403,cayssol2013floquet,rudner2013anomalous}. Driving
a topological insulator can dramatically change its nature, converting
it from a trivial to a topologically non-trivial
system~\cite{PhysRevLett.110.200403,cayssol2013floquet,rudner2013anomalous}. A number of studies in the recent past have considered both periodic
drives~\citep{mishra2015floquet,
  PhysRevLett.110.200403,eckardt2005superfluid,kitagawa2011transport}
and aperiodic
drives~\cite{nandy2017aperiodically,vcadevz2017dynamical,PhysRevB.99.155149,nandy2018steady},
leading to the discovery of many novel phenomena.
\end{comment}

\section{Introduction}  The phenomenon of many-body localization has
grabbed much attention in the last decade or
so~\cite{alet2018many,abanin2017recent,abanin2018many,basko2006metal,PhysRevLett.95.206603,nandkishore2015many}. While
disorder-induced localization (Anderson
localization~\cite{PhysRev.109.1492}) is well known, a thorough
understanding of the survival of localization in the presence of
interactions (termed as ``many-body localization
(MBL)"~\citep{PhysRev.109.1492,basko2006metal,PhysRevLett.95.206603,nandkishore2015many})
is a work in progress. Very recently, MBL-like signatures (Stark-MBL)
have been observed in a clean interacting system subjected to a static
electric
field~\citep{schulz2019stark,van2019bloch,bhakuni2019entanglement,
  taylor2019experimental}. In this article, we investigate the fate of
Stark-MBL in the presence of an additional drive.

In general, driving a many-body system heats it up as a consequence of
the energy absorption from the external drive, albeit
  slowly at high-frequencies
  ~\citep{abanin2015exponentially,d2014long,lazarides2014equilibrium,
    lazarides2014periodic,ray2018drive,ray2018signature}. Thus, the
  system approaches a prethermal phase at intermediate times followed
  by a featureless infinite-temperature-like state in the long-time
  limit~\cite{abanin2017effective,weidinger2017floquet}. Avoiding such
  a heat death scenario has been an important goal as driven systems
  can lead to many exotic features such as Floquet topological
  insulators~\cite{PhysRevLett.110.200403,cayssol2013floquet,rudner2013anomalous}
  and Floquet time-crystals~\cite{else2016floquet}. One way to avoid
this is by the inclusion of a strong disorder leading to a stable MBL
phase in the presence of high frequency
driving~\citep{ponte2015periodically,lazarides2015fate,abanin2016theory,
  d2013many,ponte2015many}, as has been observed
experimentally~\citep{bordia2017periodically} or to
  resort to the prethermal
  time-window~\cite{weidinger2017floquet,singh2019quantifying,rubio2020floquet}.
Subjecting the system to a time-periodic \emph{electric field} drive
is special as it effectively suppresses the hopping
strength~\citep{dunlap1986dynamic,dunlap1988dynamic,PhysRevB.98.045408,eckardt2009exploring},
and may be used to convert an ergodic phase into a stable MBL
phase~\cite{bairey2017driving}. The noninteracting
  limit already yields a vast variety of phenomena associated with
  electric field
  drive~\citep{dunlap1986dynamic,dunlap1988dynamic,PhysRevB.98.045408,eckardt2009exploring,holthaus1995random,holthaus1995ac,PhysRevB.99.155149,kudo2011theoretical,PhysRevB.86.075143,caetano2011wave}.
\begin{figure}
	\includegraphics[scale=0.65]{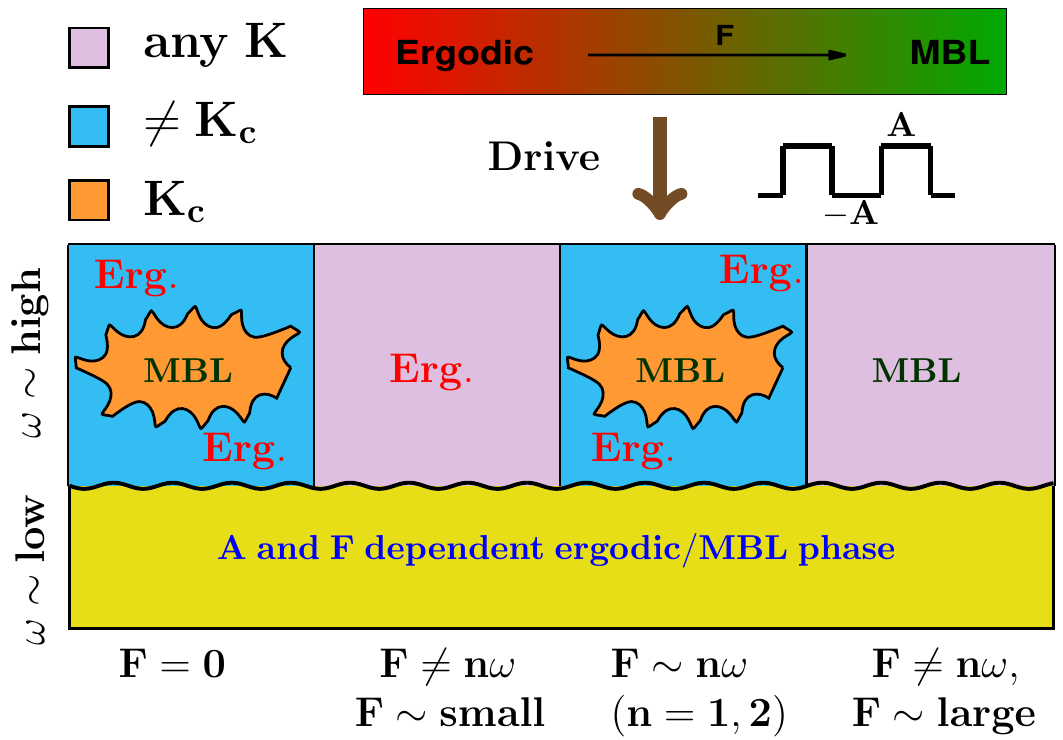}
                        \caption{ A schematic diagram of
                            the different phases based on the average
                            level spacing ratio. Top bar pertains to
                            the undriven model ($\mathcal{F}(t)=F$). For the
                            driven model, the different possibilities of the
                            ergodic and MBL phases are also  marked depending on the tuning of the parameters $K=A/\omega$ and $F$. In the high frequency, an MBL phase is obtained either by tuning the tuning the parameter at DL point or by off-resonant drive at strong field strengths. In the low frequency regime, the possibility of the ergodic and the MBL phase depends on the parameters $F$ and $A$. The actual transition from low-to-high frequency also depends on the different parameters of the system and hence shown by a wiggly line (only schematic in nature). The whole schematic is drawn while considering $K=A/\omega$ to lie below the second DL
                            point. }
		\label{schematic}
\end{figure}

An important open question has to do with the effect of a drive on a
\emph{clean} MBL system, which we address here. Specifically, we study
the model resulting from the application of an ac field comprising of
square wave pulses onto the clean MBL system of Schulz et
al~\cite{schulz2019stark}. Remarkably, the drive is found to take the undriven system
from an ergodic phase to an MBL phase and vice-versa when the
parameters are set appropriately. In the non-interacting limit, for
the case of a combined dc and square-wave driving, we obtain analytically
the conditions for dynamic localization (DL), coherent destruction of
WS localization and super-Bloch oscillations. 

Our main findings are captured schematically in Fig.~\ref{schematic}.
Keeping the dc field alone is equivalent to the undriven model, which
exhibits a phase transition from an ergodic to a Stark MBL
phase~\cite{schulz2019stark}. In the presence of a high-frequency
drive, we obtain an intricate set of possibilities dependent on how
the static electric field and the ratio of the amplitude to the
frequency ($A/\omega$) of the drive are tuned. The addition of a drive
in the zero dc field limit, induces an MBL phase if the ratio
$A/\omega$ is tuned close to the dynamic localization point, analogous
to the drive-induced MBL phase reported~\cite{bairey2017driving} in a
conventional disordered MBL model. For a large dc field, where the
undriven model yields the Stark MBL phase, the addition of resonantly
tuned drive leads to a destruction of Stark MBL for all values of the
ratio $A/\omega$ tuned away from the dynamic localization point. We
refer to this as \emph{coherent destruction of Stark-MBL}. However,
Stark MBL is found to be robust against off-resonant
drive. Importantly, the coherent destruction of
  Stark-MBL is accompanied by the appearance of a large prethermal
  window before the final infinite-temperature-like state is reached,
  as captured by the dynamics of entanglement entropy. In the
low-frequency limit, the nature of the phase obtained depends heavily
on the choice of $F$ and $A$.
% {\color{blue} All our results   are	supported by a study of the dynamics of entanglement entropy. In particular, along with providing the signatures of MBL phase, the dynamics of entanglement entropy shows a large prethermal window before the infinite-temperature fate for the set of parameters leading to coherent-destruction of Stark MBL phase. Thus, our work provides the two different mechanism of avoiding the heat-death.} 
\section{Model Hamiltonian}
The Hamiltonian of the system can be written as
\begin{eqnarray}\label{eq1}
H=-J\sum_{j=0}^{L-2}(c_{j}^{\dagger}c_{j+1}+c_{j+1}^{\dagger}c_{j})-\mathcal{F}(t)\sum_{j=0}^{L-1} j (n_{j}-\frac{1}{2})\qquad \nonumber \\ + \alpha \sum_{j=0}^{L-1} \frac{j^2}{(L-1)^2} (n_{j}-\frac{1}{2}) + V\sum_{j=0}^{L-2} (n_j-\frac{1}{2})(n_{j+1}-\frac{1}{2}),\nonumber
\\
\end{eqnarray}
 where $V$ is the nearest neighbor interaction and
  $\mathcal{F}(t) = F + A\;\text{sgn}(\sin(\omega t))$ is a combined
  dc and ac electric field, with $A$ and $\omega$ respectively being
  the amplitude and frequency of the ac field ($F_{ac}$), while $F$ is
  the dc field. The curvature term (with strength $\alpha$)
  provides a non-linearity to the dc field and thus makes
  Stark-MBL possible~\cite{schulz2019stark}. The lattice constant is
  kept at unity and natural units ($\hbar=e=1$) are adopted for all
  the calculations.
\begin{comment}
where $\mathcal{F}(t)$ is the time-dependent linear electric field,
$\alpha$ is the curvature term and $V$ is the nearest neighbor
interaction. {\color{blue} The curvature term may be thought of as a
time-independent quadratic correction to the dc part of the electric
field potential, and is included to allow the possibility of
Stark-MBL ~\cite{schulz2019stark}.} The lattice constant is kept at
unity and natural units ($\hbar=e=1$) are adopted for all the
calculations.  The time-dependent electric field is: $\mathcal{F}(t) =
F +A\;\text{sgn}(\sin(\omega t))$, where $A$ and $\omega$ respectively
are the amplitude and frequency of the ac field ($F_{ac}$), while $F$
is the static dc field.
\end{comment}

\section{Non-interacting case: Semi-classical description}
 Let us consider the non-interacting case ($V=0$), and with zero curvature
$(\alpha=0)$. The quasi-momentum can be expressed as
\begin{equation}
q_{k}(t)=k+Ft+\int_{0}^{t} d\tau
F_{ac}(\tau).
\end{equation}
Due to the dc part, the quasi-momentum is no longer a
periodic function. However, for the resonance condition
$F=n\omega$, the quasi-momentum becomes a periodic
function. Solving (as described in appendix~\ref{A1}) for the
one cycle average of quasi-energy, we get
\begin{equation}\label{8}
\epsilon(k) = -2J_\text{eff}\cos(k+\frac{n\pi}{2}),
\end{equation}
where
\begin{equation}
J_\text{eff}
=J\left\lbrace\frac{\sin(\frac{K\pi}{2} + \frac{n\pi}{2})}{(K\pi+n\pi)}+(-1)^{n}\frac{\sin(\frac{K\pi}{2}-\frac{n\pi}{2})}{(K\pi-n\pi)}\right\rbrace,
\end{equation}
and $K =A/\omega$. In the limit $F=0$, this reduces to the well known
quasi-energy dispersion for square wave driving: $\epsilon(k)=-2J
\text{sinc}\left(\frac{\pi K}{2}\right)\cos k$, where
$\text{sinc}(z)=\sin(z)/z$. The quasi-energy band collapses at the
zeros of the function $\text{sinc}(\pi K/2)$, which occurs at $K =
K_{c} = 2\nu$, $\nu\in\mathbb{Z}$, which is the condition for
dynamic localization ~\cite{eckardt2009exploring}.

For a finite $F=n\omega$, the quasi-energy spectrum can be further simplified. For even and odd $n$ respectively,
we get $\epsilon(k) = -2 J_{even} \cos k$ and $\epsilon(k) = -2 J_{odd} \sin k$, where
\begin{equation}\label{evenodd}
  J_{even}=\frac{2JK\sin(\frac{K\pi}{2})}{(K^{2}-n^2)\pi};\;\;J_{odd}=\frac{2JK\cos(\frac{K\pi}{2})}{(K^{2}-n^2)\pi}.
\end{equation}
In the even and odd cases respectively, the band collapse occurs at
$K = K_{c}=2\nu$ and $K = K_{c}=2\nu +1$, $\nu\in\mathbb{Z}$ and
$K_{c}\neq n$. At these points an initially localized wave packet
returns to its starting position. This gives the condition of
dynamic localization. For other values of $K$, and provided that
the resonance condition holds, band formation takes place and the
WS localization due to the static dc field is destroyed. A
slight detuning from resonance ($F=(n+\delta)\omega$), results in
super-Bloch oscillations with the time period given by
$T_{\text{SBO}}=\frac{2\pi}{\omega\delta}$~\citep{kudo2011theoretical,PhysRevB.86.075143,caetano2011wave}.
\begin{figure*}[t]
	\includegraphics[scale=0.75]{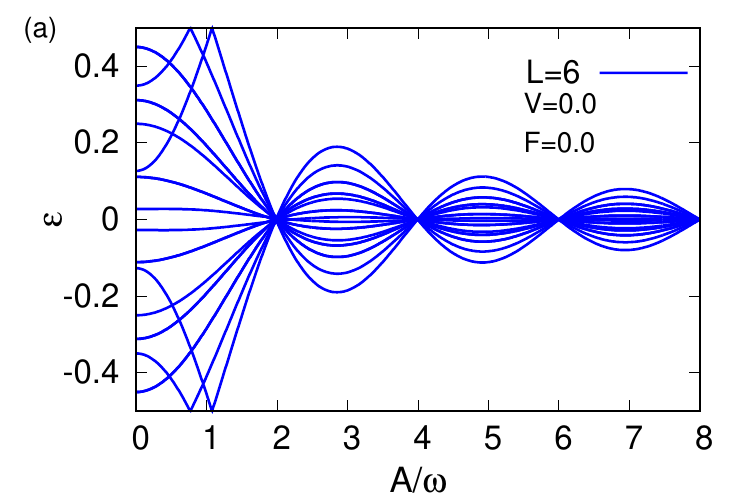}
	\includegraphics[scale=0.75]{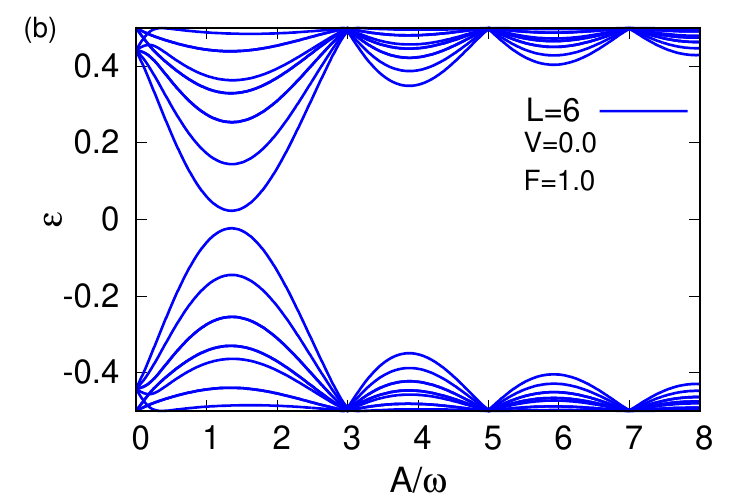}
	\includegraphics[scale=0.75]{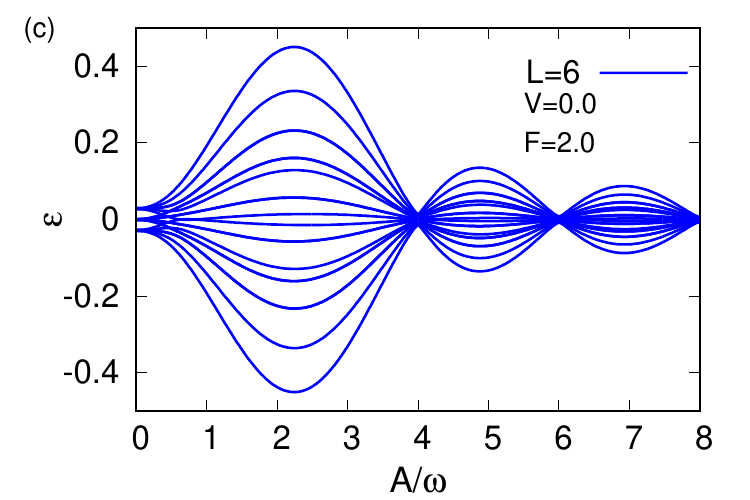}
	\includegraphics[scale=0.75]{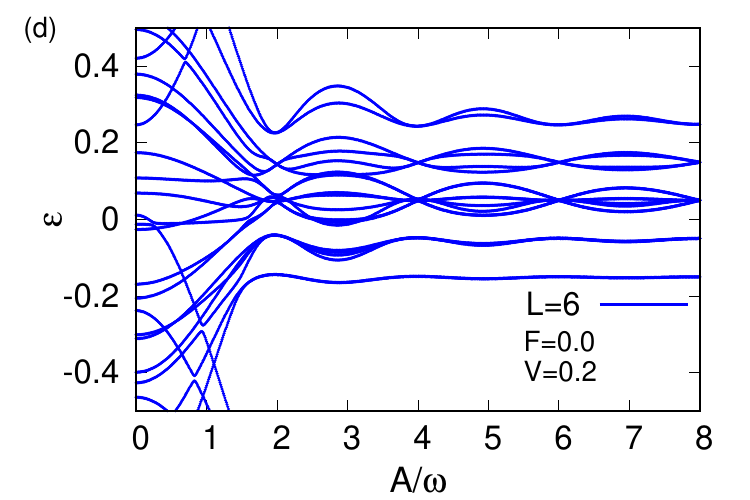}
	\includegraphics[scale=0.75]{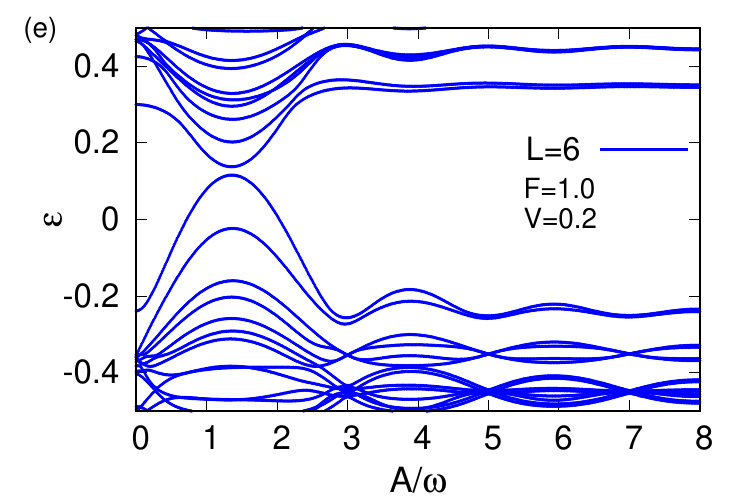}
	\includegraphics[scale=0.75]{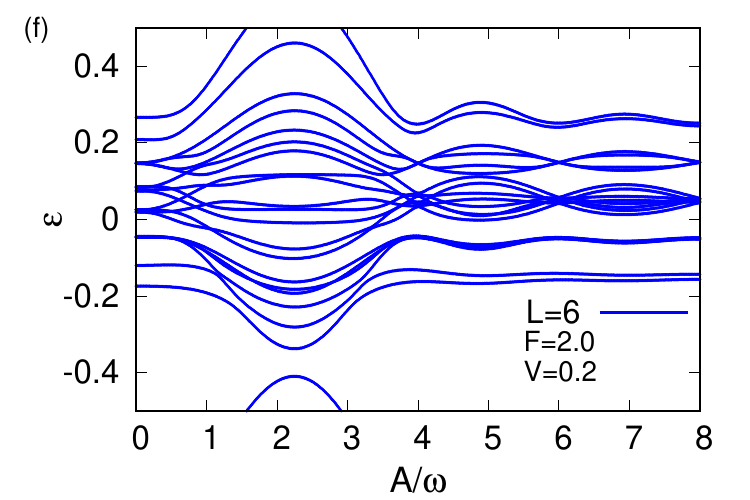}
	\caption{Top panel: Quasi-energy spectrum in the non-interacting case for different strengths of the static field. (a)
		for $F=0.0$, DL occurs for an
		even integer value of $A/\omega$. (b,c) DL for a finite value of
		the static field $F=n\omega$. The corresponding condition for DL also changes depending whether $n$ is odd (b) or even (c). Bottom panel:  Quasi-energy spectrum for the interacting case with different strengths of the static field. A finite interaction avoids the band collapse thus destabilizing DL in the presence of interactions. The other parameters are: $ L =6, \omega = 1.0, t=0.25\  \text{and filling factor =} 0.5$ in all the figures.}
	\label{qeig}
\end{figure*}

The band collapses for zero static
field and both even and odd $n$ are shown in Fig.~\ref{qeig}(a-c). The band
collapse in Fig.~\ref{qeig}(b) comes about because the quasi-energy is
conserved modulo $\omega$, and therefore the zero level is the
same as $0.5$.
\begin{figure*}
	\includegraphics[scale=0.33]{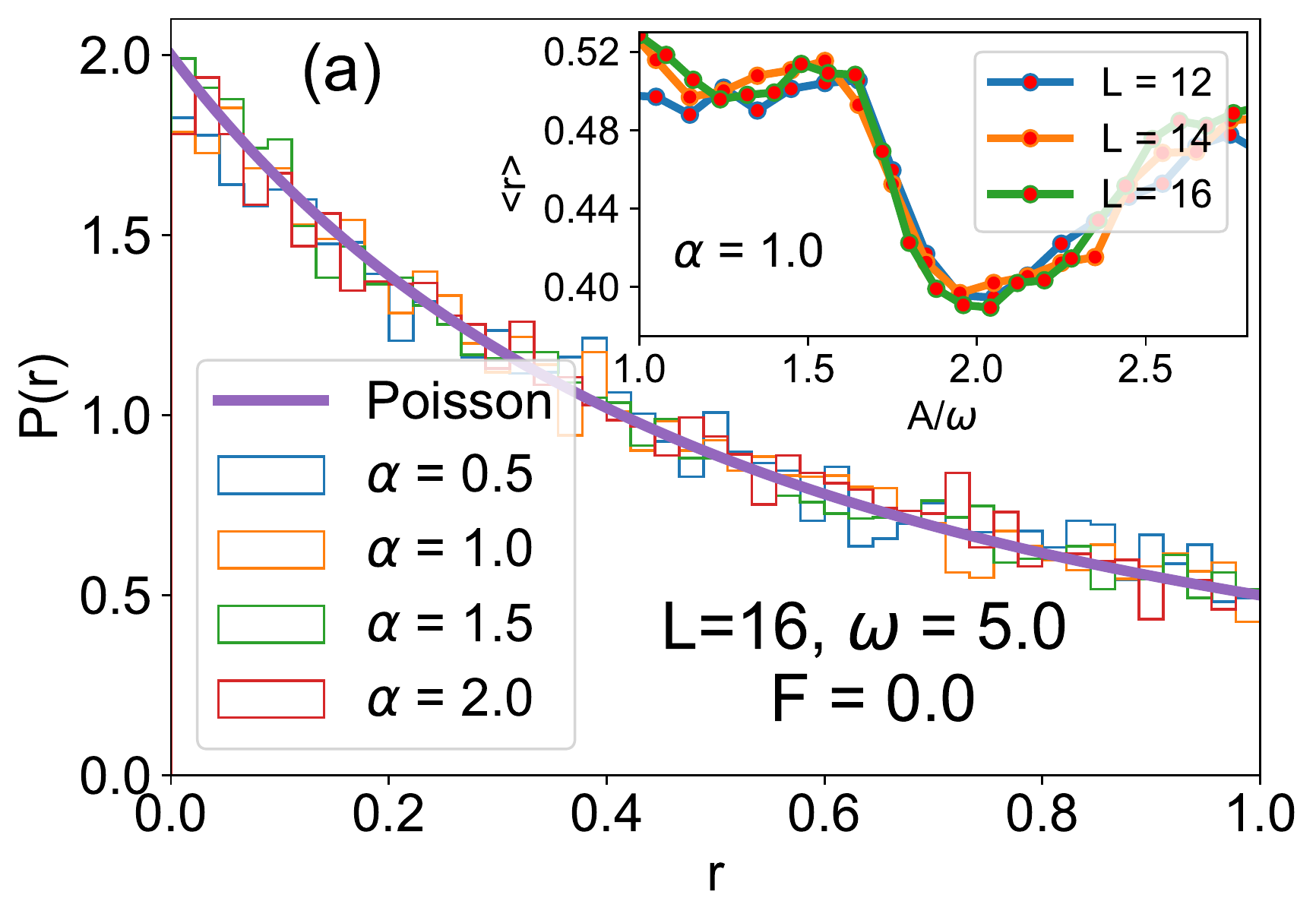}
	\includegraphics[scale=0.33]{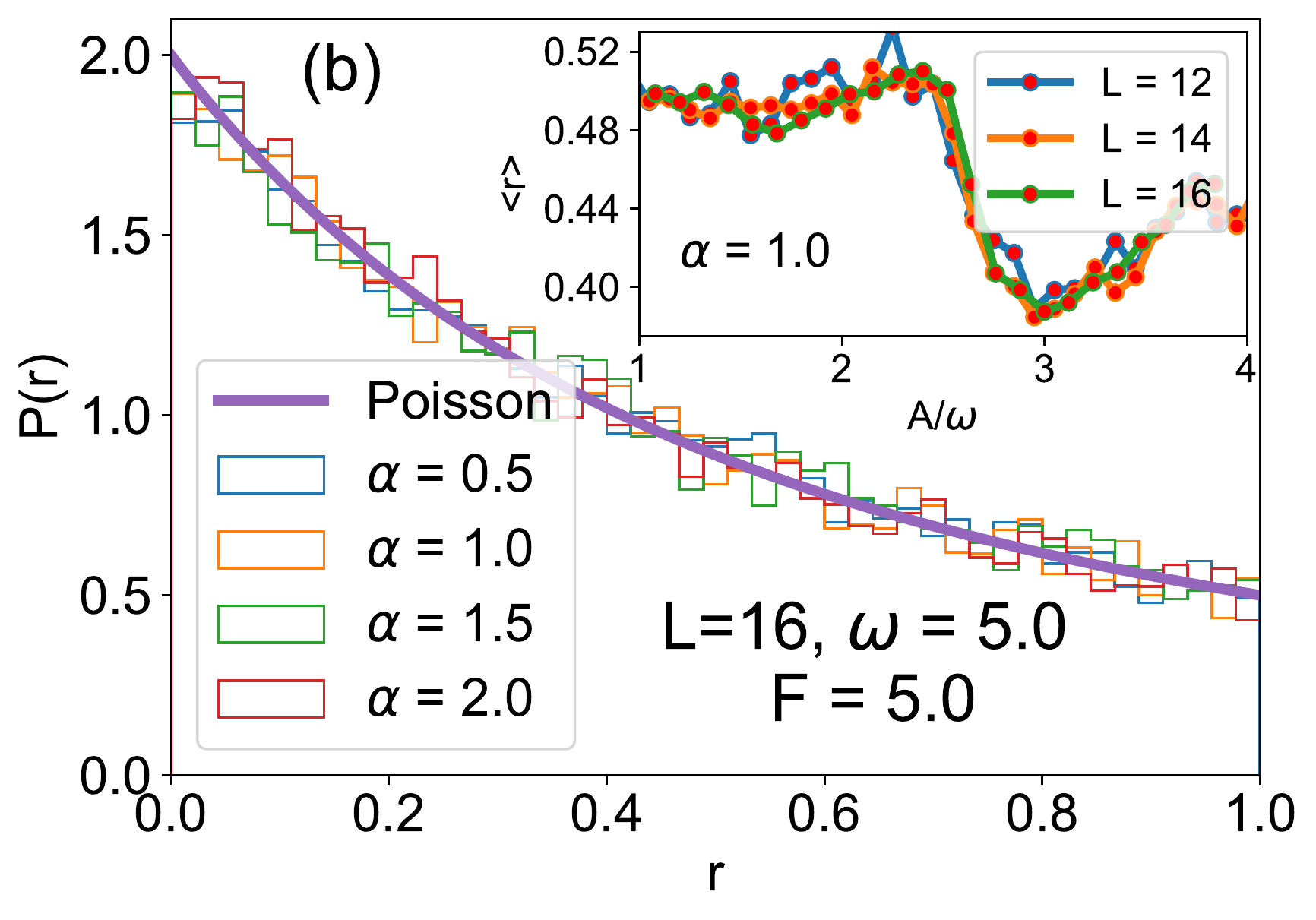}
	\includegraphics[scale=0.33]{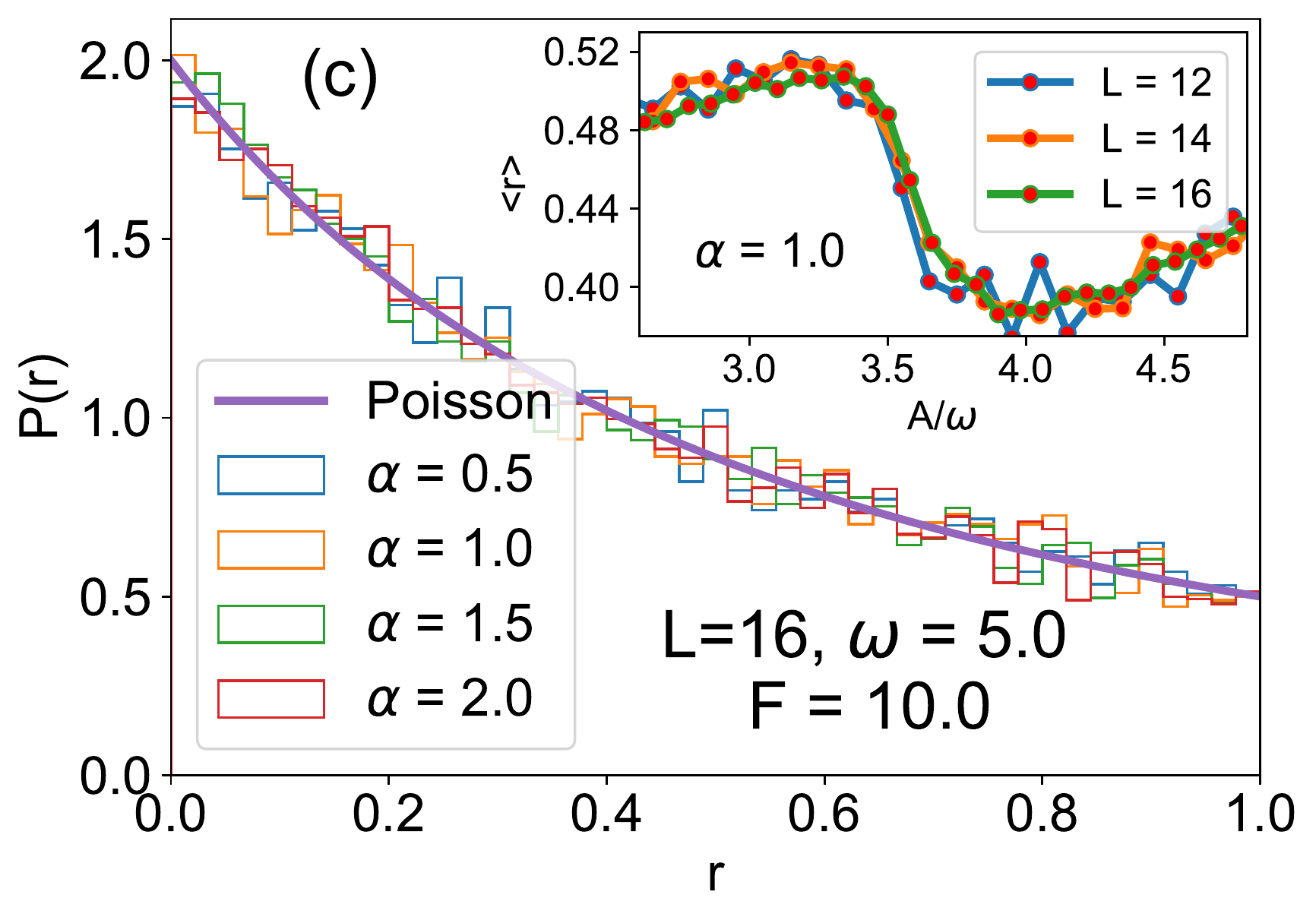}
	\includegraphics[scale=0.33]{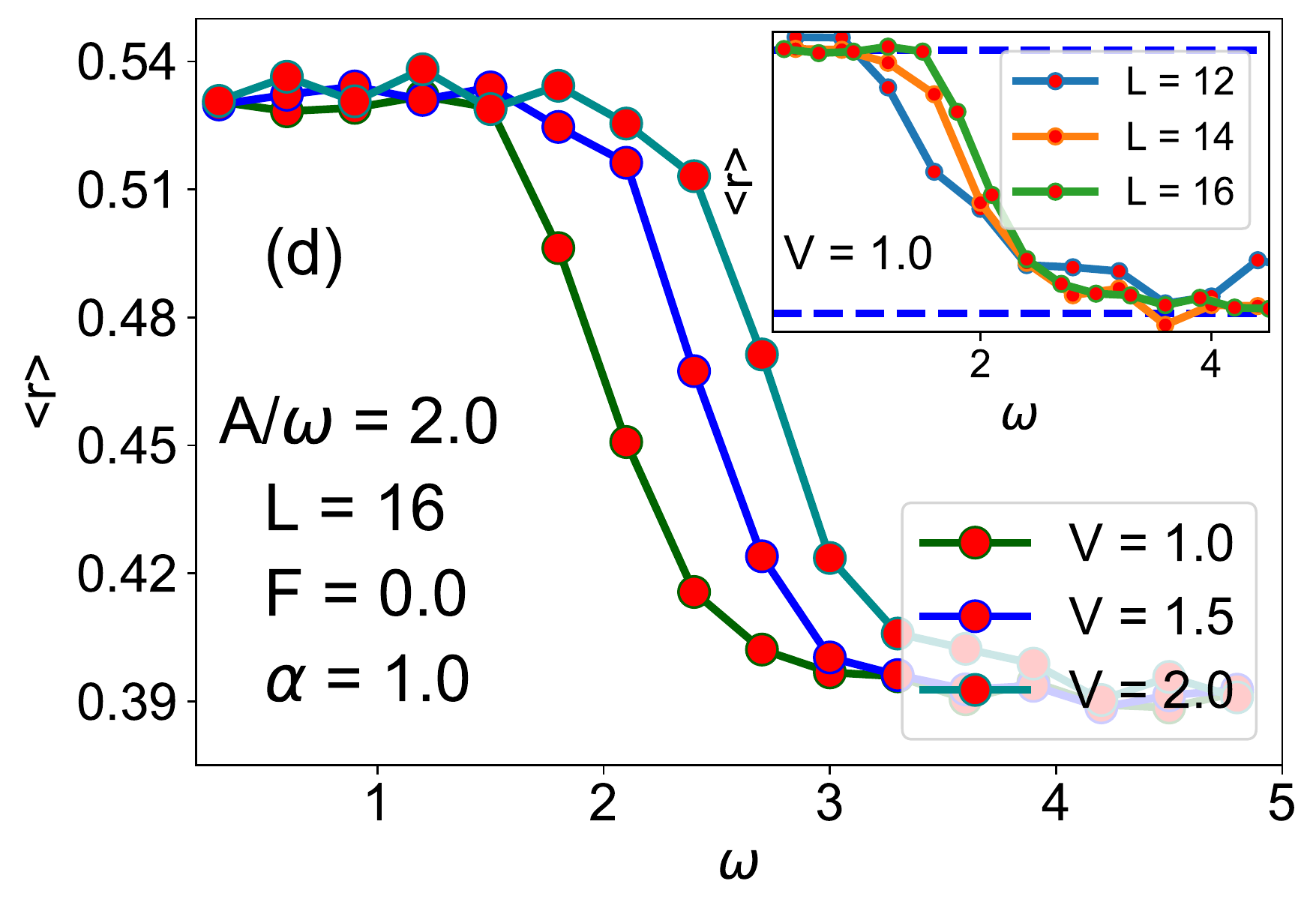}
	\includegraphics[scale=0.33]{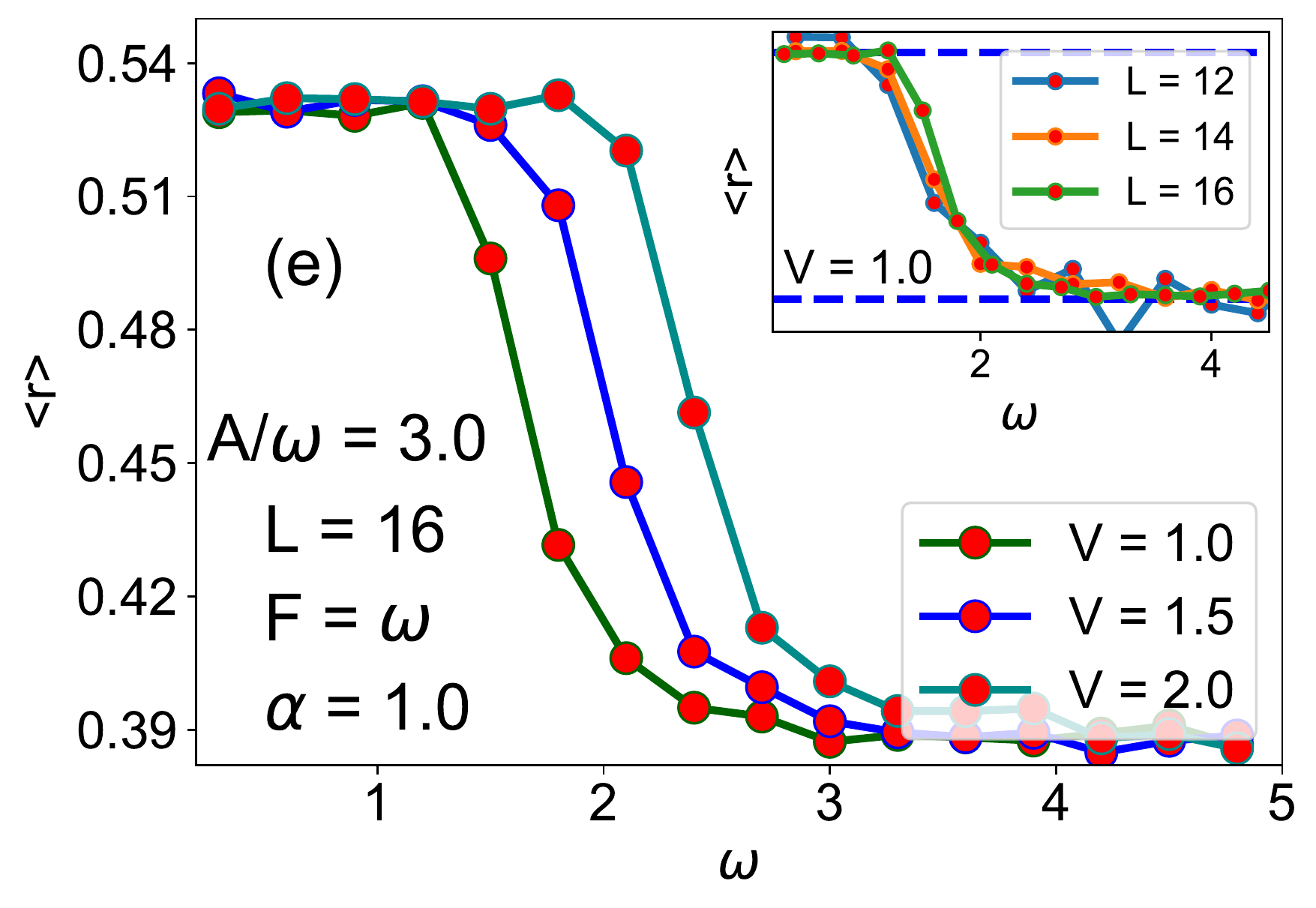}
	\includegraphics[scale=0.33]{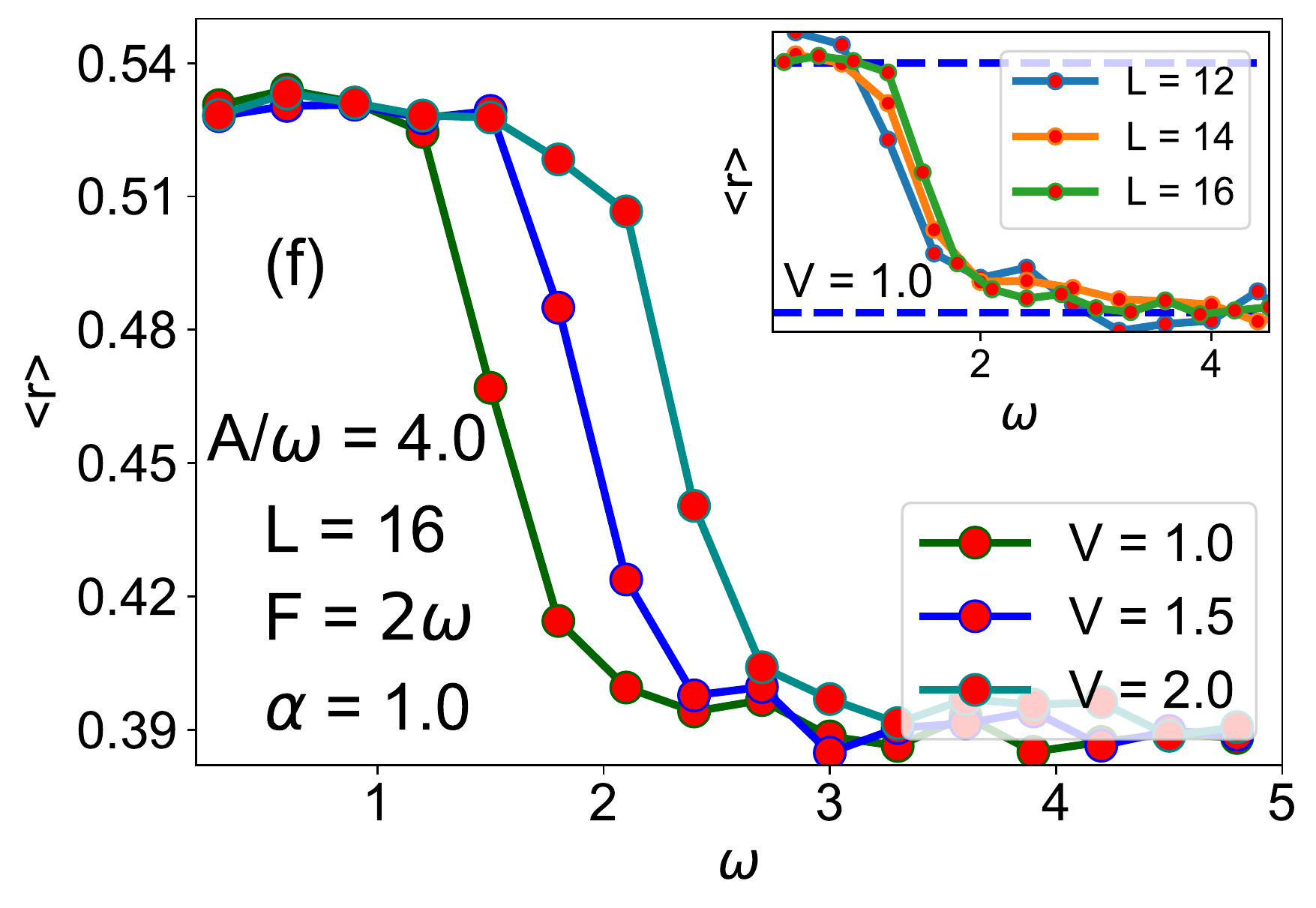}
	\caption{(a-c) Probability distribution of the quasi-energy gap ratio
		parameter for a high frequency drive ($\omega=5.0$) at the DL point. The cases $(a)$ without any dc field, $(b)$
		resonantly tuned drive ($F=n\omega$) with $n$-odd and $(c)$
		with $n$-even, are shown. The insets in (a-c)
		shows the average level-spacing ratio as a function of the ratio
		$A/\omega$ for different system sizes. % In all the cases, the MBL phase is found by tuning the ratio at the dynamic localization point, whereas away from the dynamic localization point an ergodic phase is observed.
		(d-f) Frequency dependence of the average
			level-spacing ratio for various interaction strengths. In all the cases, the ratio $A/\omega$ is tuned
			at the DL point. (d) For zero dc field case,
			%the data suggest a critical frequency around $\omega_c \approx 2$. 
			(e,f) For resonant drive with (e) $n$ odd and (f) $n$ even. %In both the cases, the critical value seems to be less than $2$. 
			The insets in (d-f) show the finite-size scaling of the frequency dependence transition for $\alpha=1.0$. The other parameters are: $J=1.0,
			V=1.0$.}
	\label{ravqeig}
\end{figure*}
\section{Interacting model}
In this section, we discuss how the presence of many-body interactions affect the dynamic localization and how the MBL phase can be obtained by including the curvature term.
\subsection{Absence of dynamic localization} 
For any general time-periodic Hamiltonian,
the Floquet operator over one cycle can be expressed in terms of
Floquet Hamiltonian $H_F$ as: $U(T) =\mathcal{T}\int_{0}^{T}
e^{-iH(t)} dt = e^{-i H_F T}$, where $\mathcal{T}$ represents the time
ordering and $T$ is the time period of the drive.  For the square wave drive, defining $H_{+}$
and $H_{-}$ as the Hamiltonians for the first and second half of the
driving period respectively, the Floquet operator can be simplified to
$U(T) = (e^{-iH_{-}T/2}e^{-iH_{+}T/2})$. The required quasi-energies and the
Floquet eigenstates are then calculated by numerically diagonalizing
the Floquet operator (upto $L=16$ at half-filling). 

The obtained quasi-energy spectrum for the interacting case with
$\alpha = 0$, is plotted in Fig.~\ref{qeig}(d-f). We
observe that the quasi-energy spectrum for all the cases of
$F=0$,$\ n$-odd, and $n$-even, has a tendency to avoid the band
collapse in contrast to the non-interacting problem (Fig.~\ref{qeig}(a-c)) where the band does collapse at certain special points. This signifies the destruction of dynamic localization in the presence of many-body interactions.

\subsection{MBL in the presence of a curvature term}
 Although interactions are inimical to dynamic localization, the presence of a
non-zero curvature term can lead to the MBL phase.  As
  evident from the non-interacting case with $\alpha = 0$, the drive
  re-normalizes the hopping strength (Eq.~\ref{evenodd}) and in the high-frequency limit, the Hamiltonian can be effectively written as a nearest-neighbor
  hopping model with re-normalized hopping strength for resonant
  driving. In the presence of interactions this picture fails as the
  quasi-momentum is no longer conserved. Nevertheless, there will be
  some residual suppression of the hopping strength ~\cite{luitz2017absence,bairey2017driving}.  With only interactions, and tuning the parameters at the dynamic localization point ($J_{\text{eff}} = 0$), this leads
  to the destruction of dynamic localization~\cite{luitz2017absence} and this eventually leads to a de-localization effect, however we point out that an additional non-zero onsite potential (curvature term) yields MBL around the dynamic localization point.
  %as evident from the following effective Hamiltonian~\cite{luitz2017absence,bairey2017driving}.
  %\begin{equation}
  %\tilde{H} = \sum_{n\ne m} J_0 h_{nm}c_{n}^{\dagger}c_{m} + g(\{n_{i}\}) +  \sum_{n\ne m} h_{nm}c_{n}^{\dagger}c_{m}\sum_{s}J_s e^{i\omega st},
  %\end{equation}
  %where $J_0$ corresponds to the re-normalized hopping term obtained by
  %calculating the time-averaged Hamiltonian.

\subsubsection{High-frequency driving}
To characterize the MBL phase, we study the probability distribution of the level spacing. Fig.~\ref{ravqeig} shows the probability distribution of the
quasi-energy gap-ratio parameter: $r_n =
\text{min}(\delta_n/\delta_{n+1},\delta_{n+1}/\delta_{n})$, where
$\delta_n$ is the difference between the $n^{th}$ and $(n-1)^{th}$
quasi-energy levels, for a system of size $L=16$ and various values of
the curvature term for a large driving frequency $\omega = 5$. For all
the cases $(F=0, n\omega)$, the probability distribution agrees with
the Poisson distribution: $P(r) = 2/(1+r)^2$ and suggests an MBL phase
at these special points. The inset in Fig.~\ref{ravqeig}(a) shows the
level-spacing ratio as a function of $A/\omega$ for zero dc
field. Although the un-driven model $F=0$, is in the ergodic phase~\cite{schulz2019stark}, the application
of drive leads to the MBL phase with a proper tuning of the ratio
$A/\omega$ to the dynamic localization point ($A/\omega = 2\nu$ with
$\nu\in\mathbb{Z}$) of the non-interacting problem.

For the case where an additional static
field is also present and satisfies the resonance condition:
$F=n\omega$, the condition for dynamic localization in the
non-interacting limit depends on whether the integer $n$ is odd or
even (Eq.~\ref{evenodd}).  We therefore explore both the cases setting
$F=5,10$ (corresponding to $n=1,2$ respectively) as shown in the
insets of the Fig.~\ref{ravqeig}(b,c). The un-driven model
corresponding to these field strengths shows the Stark-MBL phase. By
turning on the drive in this case, we find that the Stark-MBL phase
can be destroyed with \emph{resonant} driving. We therefore term this
phenomenon as ``coherent destruction of Stark-MBL". However, close to the points of dynamic localization,
Stark-MBL is found to remain intact.  Thus we see that by a proper
tuning of $F$, $A$, and $\omega$ it is possible to convert the ergodic
phase into the MBL phase and vice versa.

  We have seen that tuning the ratio $A/\omega$ at the dynamic
	localization point, seems to favour MBL. However, this turns out to be
	true only for sufficiently high frequencies, and in fact, there is a
	frequency-dependent transition from the ergodic to the MBL phase. To
	demonstrate this, we tune the ratio $A/\omega$ at the dynamic
	localization point and vary the driving frequency $\omega$, while
	keeping the ratio $A/\omega$ fixed. For the zero static field case,
	Fig~\ref{ravqeig} (d) shows the average level spacing for various interaction strengths. Fig~\ref{ravqeig} (e) and
	Fig~\ref{ravqeig} (f) consider the scenario when an additional dc
	field is present. Here, in addition to keeping the ratio $A/\omega$
	fixed at the dynamic localization point, we also keep the ratio
	$F/\omega$ fixed (at $1$,$2$ respectively) as $\omega$ is varied. It can be seen that only at sufficiently high frequency an MBL phase is observed. Moreover, the frequency  required to observe the MBL phase also increases on increasing the interaction strength. This can be understood from the
	phase diagram of the undriven system (see appendix~\ref{A2}), where
	the ergodic-to-MBL phase transition occurs at a higher value of the
	field strength on increasing the interaction strength. In the
	presence of the drive, an MBL phase is obtained only when the
	effective field in the two half-cycles ($F\pm A$) lies in the
	deep-MBL region of the undriven system where the drive only mixes
	the localized states of the undriven system. % The data for all the system sizes cross at $\omega \approx 2$, which suggests a critical frequency of $\omega_c \approx 2$, for the chosen parameters here.
	%We see that the critical frequency is now found to be smaller than with the zero dc field.
	The insets of Fig~\ref{ravqeig} (d-f) show the finite-size scaling for $V=1.0$ and supports the argument presented above. 
\begin{figure*}[t]
	\includegraphics[scale=0.52]{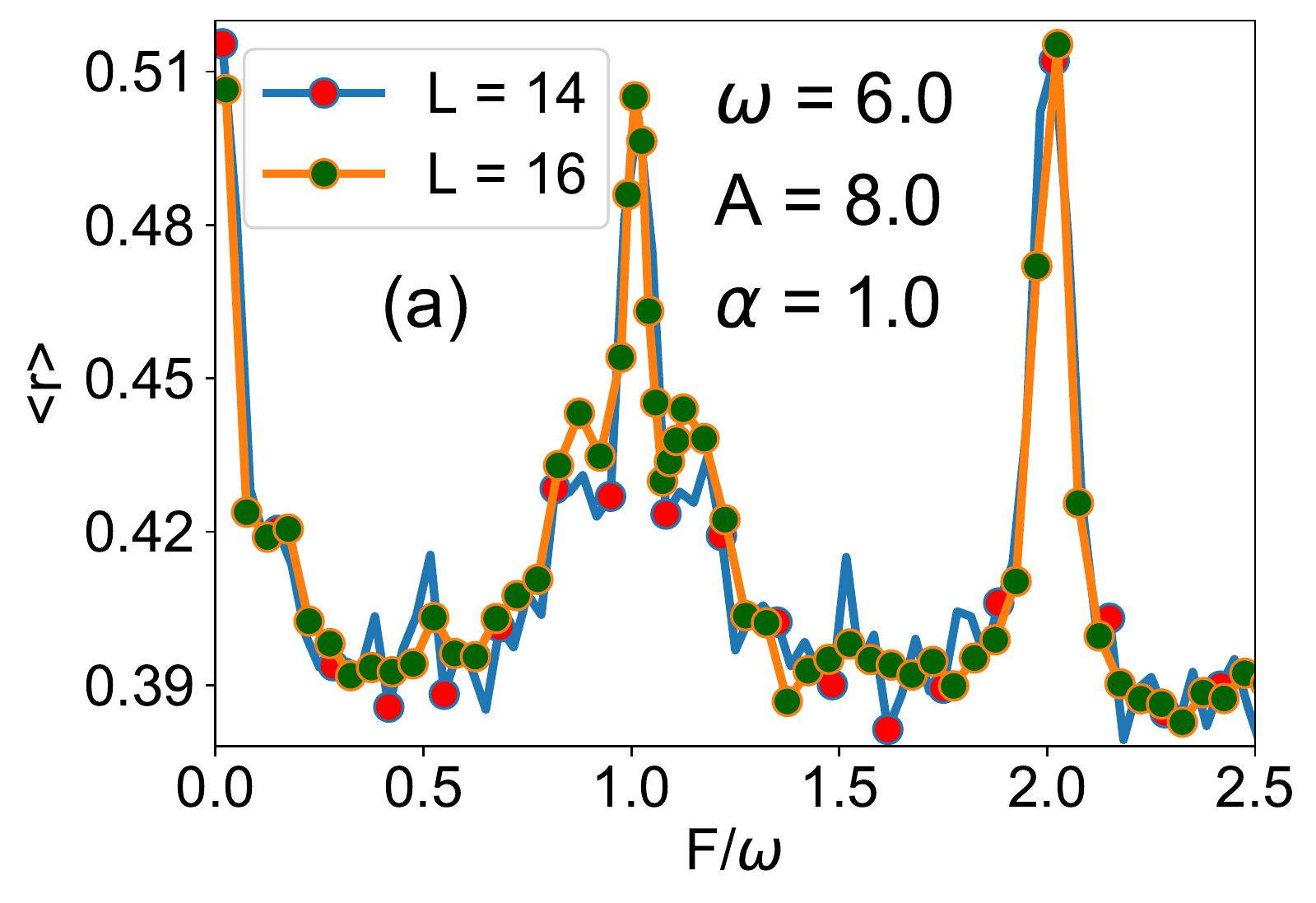}\hspace{10pt}
	\includegraphics[scale=0.52]{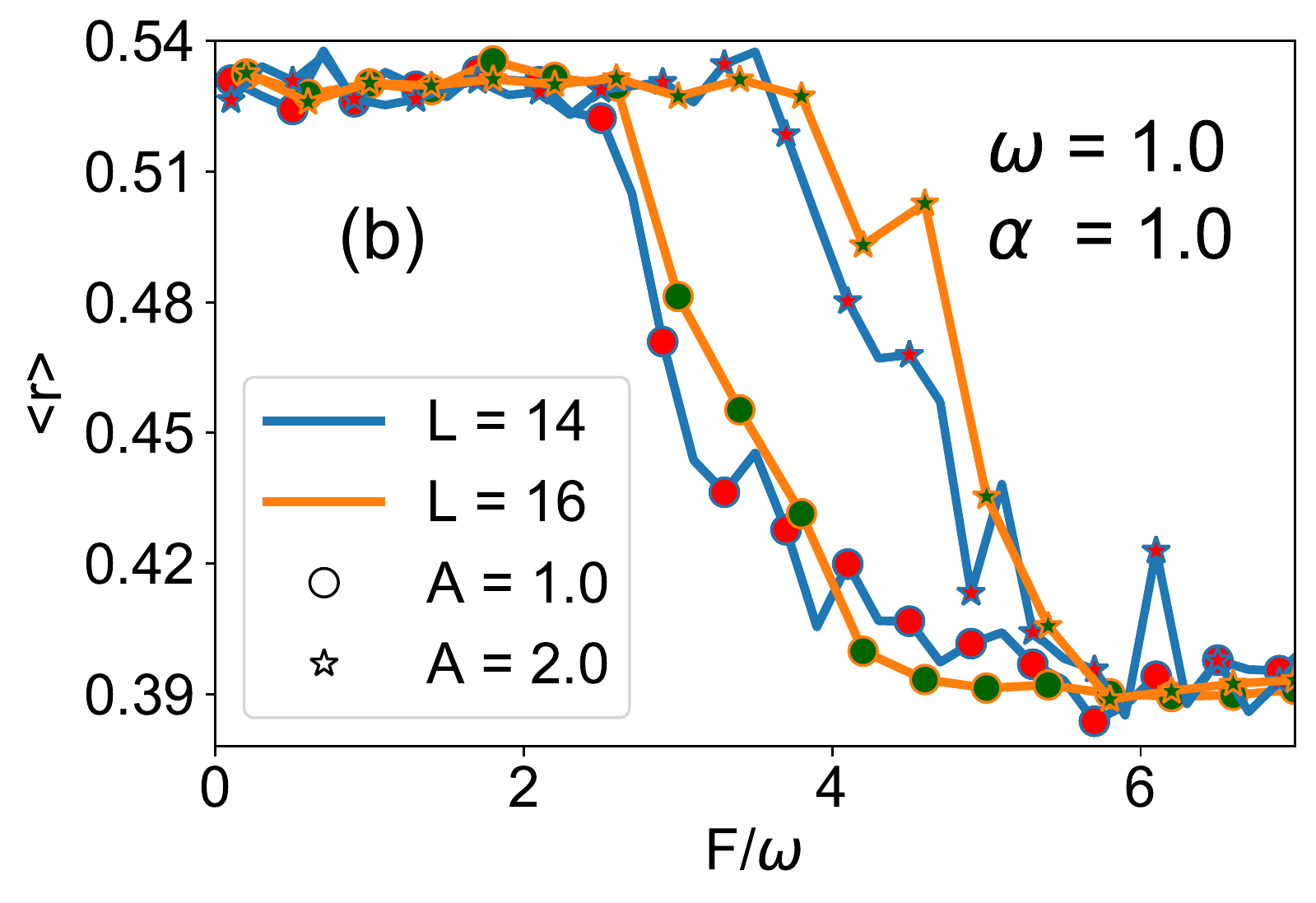}
	\caption{(a) High Frequency drive: Average level-spacing ratio as a function of $F/\omega$ and away from the DL point ($\omega = 6.0, A=8.0$). For the condition
		$F=n\omega$, Stark-MBL is destroyed while for off-resonant driving,
		the Stark-MBL phase is robust. (b) Low frequency drive: The average level spacing ratio as a function of $F/\omega$ for $A=1.0, 2.0$ for $(\omega = 1.0)$. The transition from ergodic to the MBL phase is found to be dependent on the parameters $A$ and $F$. The other parameters are: $J=1.0,V=1.0$, and $\alpha=1.0$}
	\label{ravoffreso}
\end{figure*}
\begin{figure*}[t]
	\includegraphics[scale=0.36]{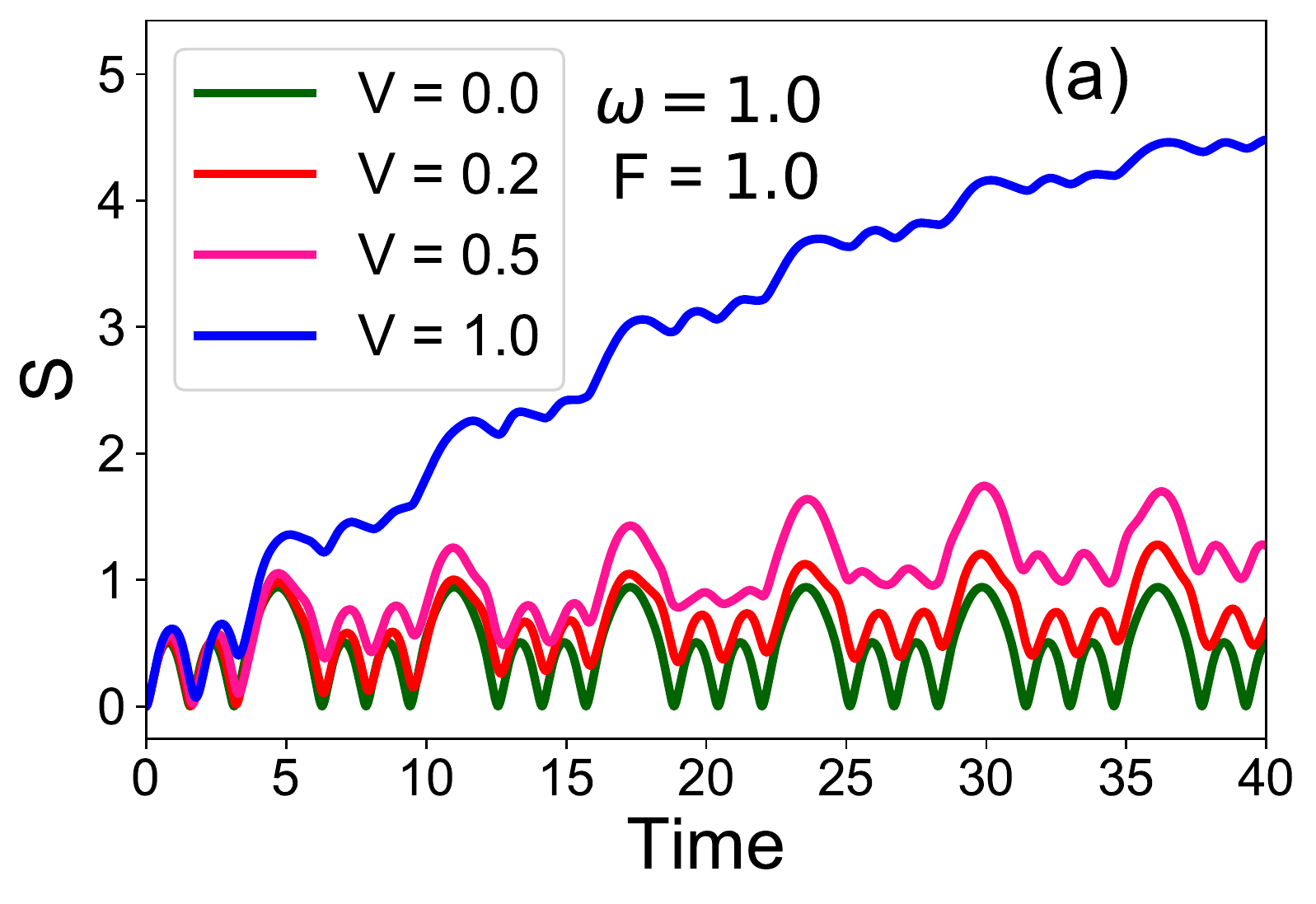}
	\includegraphics[scale=0.36]{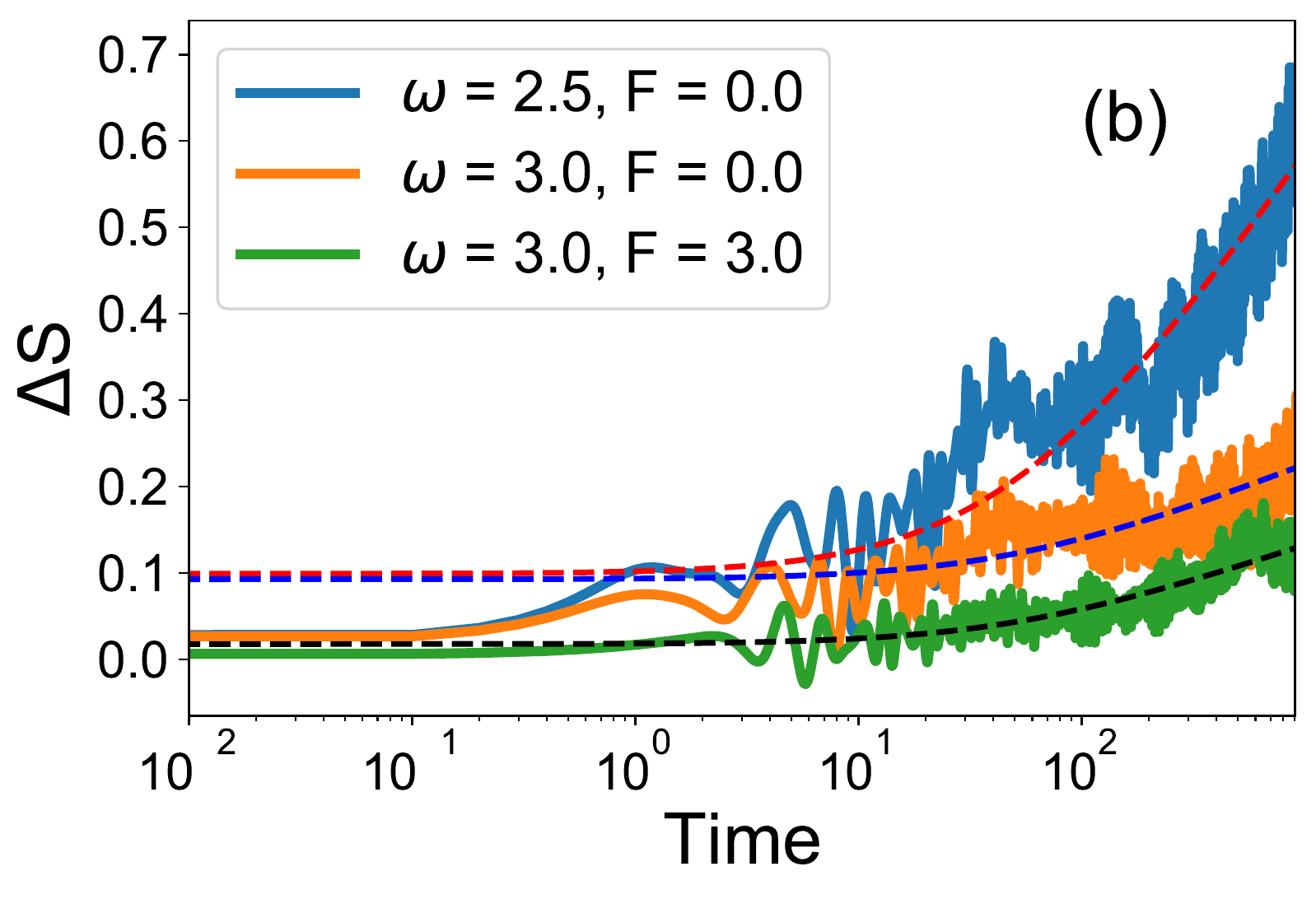}
	\includegraphics[scale=0.36]{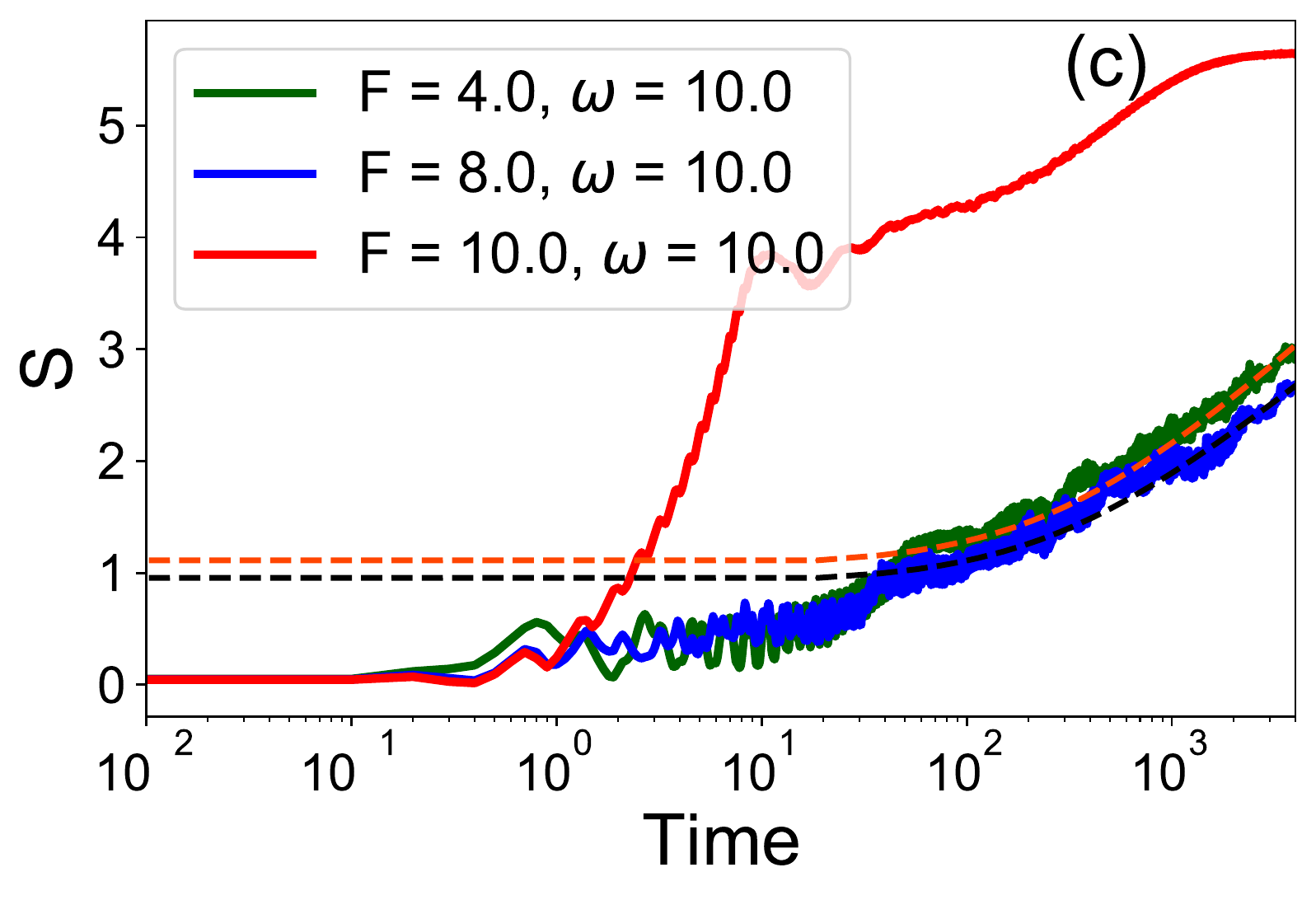}
	\caption{(a) Absence of DL in the presence of
		many-body interactions (with $\alpha = 0$). A departure from perfect periodicity is
		observed. (b) Dynamics of the difference in EE
		($\Delta S$) between the interacting and the corresponding
		non-interacting limit, at the DL point. The plots are smoothed
		out by convolution with a Gaussian: $w(n)=e^{-(n/\sigma)^2/2}$, with
		$\sigma=5$. (c) Coherent destruction of Stark-MBL phase at
		resonant driving and the robustness of Stark-MBL phase at
		off-resonant driving from the EE dynamics. The
		fitting is done with the curve: $f(x)=a \log(x)+b$. The other
		parameters are: $L=18,J=1.0,V=1.0, \alpha = 1.0$.
	}
	\label{entmbl}
\end{figure*} 

Finally, we consider the case of an off-resonant
drive.  Fig.~\ref{ravoffreso}(a) shows the
level-spacing ratio as a function of the ratio $F/\omega$. Only when
the ratio $F/\omega$ approaches an integer, the level spacing ratio
satisfies the Wigner-Dyson statistics signifying an ergodic phase for
these special ratios. Thus, we see that in the high-frequency regime,
Stark-MBL remains robust against off-resonant drive. 
\subsubsection{Low-frequency driving}
In the low-frequency regime, the nature of the 
phase crucially depends on the choice of the driving amplitude and the
static field strength. Fig.~\ref{ravoffreso}(b) shows the variation of the average
level spacing ratio as a function of the static dc field for different
driving amplitudes. It can be seen that the ergodic-to-MBL phase transition depends on the choice of driving amplitude. The driven model yields an MBL
phase when both $F+A$ and $F-A$ lie in the MBL phase of the undriven
model, where the drive only mixes the localized eigenstates of
the undriven system. On the other hand, when one or both of the
parameters: $F+A$ or $F-A$ falls into the ergodic region of the
undriven model, an ergodic phase is observed. We infer that if the
drive mixes both localized and extended states, it is extended states
that dominate.

\section{ Dynamics of entanglement entropy}
For a system in a pure
state, the entanglement entropy (EE) of a subsystem $A$ is defined as:
$S_{A} = -\text{Tr}(\rho_{A}\text{ln}\rho_{A})$, where $\rho_{A}$ is
the reduced density matrix of the subsystem $A$ obtained by tracing
out the degrees of freedom of the other subsystem $B$. To study the
dynamics of the entanglement entropy, we start with an initial product
state (where all the particles occupy the even sites) and use an exact
numerical approach based on the re-orthogonalized Lanczos
algorithm~\cite{van2019bloch,luitz2017ergodic} for the time
evolution. Due to the interactions in the MBL phase, a
logarithmic growth of the entanglement entropy is
expected~\citep{vznidarivc2008many,PhysRevLett.109.017202,PhysRevLett.110.260601}.

We first consider the limit $\alpha = 0$, and study the stability of
dynamic localization in the presence of interactions. The dynamics of
entanglement entropy for various interaction strengths is plotted in
Fig.~\ref{entmbl}(a), where the ratio $A/\omega$ is tuned at the
dynamic localization point.  The entanglement entropy here starts to
grow in time as opposed to the non-interacting case where as a
consequence of the band collapse, it shows an oscillatory behavior and
recurs to its initial value %(zero due to the choice of the initial
state) at times $t=mT$ with $m$ being a positive integer.

We now turn to the case with a finite curvature strength ($\alpha =
1.0$), where an MBL phase is found at sufficiently high
frequencies.% Here, we investigate the stability of the MBL phase from a dynamical perspective. 
We first consider the case where the MBL
phase is obtained by tuning the ratio $A/\omega$ at the dynamic
localization point. We define the quantity: $\Delta S = S(t,V)
-S(t,V=0)$ as the difference between the entanglement entropy of the
interacting and the corresponding non-interacting
limit. Fig.~\ref{entmbl}(b) shows the dynamics of $\Delta S$ as a
function of time for different sets of the frequency and the static
field. In all the cases, a logarithmic behavior is observed which
signifies an MBL-like phase at these points.

The dynamics of the entanglement entropy for the parameters tuned away
from the dynamic localization point is shown in
Fig.~\ref{entmbl}(c). It can be seen that for resonant drive
($F=n\omega$), the Stark-MBL phase is destroyed. The entanglement
entropy increases rapidly for smaller times followed by a
slow growth for the intermediate times and finally saturates to its
thermal value. This slow growth of entanglement entropy in the
intermediate times is a signature of \emph{Floquet
  prethermalization}~\citep{abanin2017effective,mori2016rigorous,abanin2017rigorous,
  luitz2019prethermalization}, where the system prethermalizes before
reaching an infinite-temperature-like state at high
frequencies. For the off-resonant drive at high frequency, the
entanglement entropy shows the usual logarithmic growth signifying the
robustness of the Stark-MBL phase. 
 
\section{Summary and Conclusions}.  To summarize, we study a clean
interacting system driven by a combined ac and dc electric field. The
underlying non-interacting problem is itself of interest, and we
semi-classically obtain the condition for dynamic localization,
coherent destruction of WS localization and super Bloch
oscillations. In the presence of interactions, generic clean many-body
systems under a drive, reach a featureless infinite-temperature-like
state. In contrast, we find that our system can avoid such `heat death', under
high-frequency drive.  This is achieved either by tuning the system at
the dynamic localization point of the corresponding noninteracting
model, or by subjecting the system to off-resonant drive.

We further study the fate of the Stark-MBL phase in the presence of an
additional drive. Observing that the effects of low-frequency driving
are heavily dependent on the field strength and the amplitude of the
drive, we focus on high-frequency driving, uncovering an intricate set
of possible phases. One striking possibility is that of generating an
MBL phase from the undriven ergodic phase by the application of a pure
ac field. A second remarkable possibility appears for sufficiently
large dc field, where it is possible to destroy Stark-MBL, by the
application of a resonantly tuned drive provided that the ratio
$A/\omega$ is tuned away from the dynamic localization point. We
term this as `coherent destruction of Stark-MBL'. On the other hand,
the Stark-MBL phase is found to be robust against off-resonant drive.

\section*{Acknowledgment}
We are grateful to the High Performance Computing(HPC) facility at
IISER Bhopal, where large-scale calculations in this project were run.
A.S is grateful to SERB for the grant (File Number: CRG/2019/003447),
and the DST-INSPIRE Faculty Award [DST/INSPIRE/04/2014/002461]. D.S.B
acknowledges PhD fellowship support from UGC India.
\bibliography{ref}

\begin{thebibliography}{50}%
\makeatletter
\providecommand \@ifxundefined [1]{%
 \@ifx{#1\undefined}
}%
\providecommand \@ifnum [1]{%
 \ifnum #1\expandafter \@firstoftwo
 \else \expandafter \@secondoftwo
 \fi
}%
\providecommand \@ifx [1]{%
 \ifx #1\expandafter \@firstoftwo
 \else \expandafter \@secondoftwo
 \fi
}%
\providecommand \natexlab [1]{#1}%
\providecommand \enquote  [1]{``#1''}%
\providecommand \bibnamefont  [1]{#1}%
\providecommand \bibfnamefont [1]{#1}%
\providecommand \citenamefont [1]{#1}%
\providecommand \href@noop [0]{\@secondoftwo}%
\providecommand \href [0]{\begingroup \@sanitize@url \@href}%
\providecommand \@href[1]{\@@startlink{#1}\@@href}%
\providecommand \@@href[1]{\endgroup#1\@@endlink}%
\providecommand \@sanitize@url [0]{\catcode `\\12\catcode `\$12\catcode
  `\&12\catcode `\#12\catcode `\^12\catcode `\_12\catcode `\%12\relax}%
\providecommand \@@startlink[1]{}%
\providecommand \@@endlink[0]{}%
\providecommand \url  [0]{\begingroup\@sanitize@url \@url }%
\providecommand \@url [1]{\endgroup\@href {#1}{\urlprefix }}%
\providecommand \urlprefix  [0]{URL }%
\providecommand \Eprint [0]{\href }%
\providecommand \doibase [0]{http://dx.doi.org/}%
\providecommand \selectlanguage [0]{\@gobble}%
\providecommand \bibinfo  [0]{\@secondoftwo}%
\providecommand \bibfield  [0]{\@secondoftwo}%
\providecommand \translation [1]{[#1]}%
\providecommand \BibitemOpen [0]{}%
\providecommand \bibitemStop [0]{}%
\providecommand \bibitemNoStop [0]{.\EOS\space}%
\providecommand \EOS [0]{\spacefactor3000\relax}%
\providecommand \BibitemShut  [1]{\csname bibitem#1\endcsname}%
\let\auto@bib@innerbib\@empty
%</preamble>
\bibitem [{\citenamefont {Alet}\ and\ \citenamefont
  {Laflorencie}(2018)}]{alet2018many}%
  \BibitemOpen
  \bibfield  {author} {\bibinfo {author} {\bibfnamefont {F.}~\bibnamefont
  {Alet}}\ and\ \bibinfo {author} {\bibfnamefont {N.}~\bibnamefont
  {Laflorencie}},\ }\href@noop {} {\bibfield  {journal} {\bibinfo  {journal}
  {Comptes Rendus Physique}\ }\textbf {\bibinfo {volume} {19}},\ \bibinfo
  {pages} {498} (\bibinfo {year} {2018})}\BibitemShut {NoStop}%
\bibitem [{\citenamefont {Abanin}\ and\ \citenamefont
  {Papi{\'c}}(2017)}]{abanin2017recent}%
  \BibitemOpen
  \bibfield  {author} {\bibinfo {author} {\bibfnamefont {D.~A.}\ \bibnamefont
  {Abanin}}\ and\ \bibinfo {author} {\bibfnamefont {Z.}~\bibnamefont
  {Papi{\'c}}},\ }\href@noop {} {\bibfield  {journal} {\bibinfo  {journal}
  {Annalen der Physik}\ }\textbf {\bibinfo {volume} {529}},\ \bibinfo {pages}
  {1700169} (\bibinfo {year} {2017})}\BibitemShut {NoStop}%
\bibitem [{\citenamefont {Abanin}\ \emph {et~al.}(2018)\citenamefont {Abanin},
  \citenamefont {Altman}, \citenamefont {Bloch},\ and\ \citenamefont
  {Serbyn}}]{abanin2018many}%
  \BibitemOpen
  \bibfield  {author} {\bibinfo {author} {\bibfnamefont {D.~A.}\ \bibnamefont
  {Abanin}}, \bibinfo {author} {\bibfnamefont {E.}~\bibnamefont {Altman}},
  \bibinfo {author} {\bibfnamefont {I.}~\bibnamefont {Bloch}}, \ and\ \bibinfo
  {author} {\bibfnamefont {M.}~\bibnamefont {Serbyn}},\ }\href@noop {}
  {\bibfield  {journal} {\bibinfo  {journal} {arXiv preprint arXiv:1804.11065}\
  } (\bibinfo {year} {2018})}\BibitemShut {NoStop}%
\bibitem [{\citenamefont {Basko}\ \emph {et~al.}(2006)\citenamefont {Basko},
  \citenamefont {Aleiner},\ and\ \citenamefont {Altshuler}}]{basko2006metal}%
  \BibitemOpen
  \bibfield  {author} {\bibinfo {author} {\bibfnamefont {D.~M.}\ \bibnamefont
  {Basko}}, \bibinfo {author} {\bibfnamefont {I.~L.}\ \bibnamefont {Aleiner}},
  \ and\ \bibinfo {author} {\bibfnamefont {B.~L.}\ \bibnamefont {Altshuler}},\
  }\href@noop {} {\bibfield  {journal} {\bibinfo  {journal} {Annals of
  physics}\ }\textbf {\bibinfo {volume} {321}},\ \bibinfo {pages} {1126}
  (\bibinfo {year} {2006})}\BibitemShut {NoStop}%
\bibitem [{\citenamefont {Gornyi}\ \emph {et~al.}(2005)\citenamefont {Gornyi},
  \citenamefont {Mirlin},\ and\ \citenamefont
  {Polyakov}}]{PhysRevLett.95.206603}%
  \BibitemOpen
  \bibfield  {author} {\bibinfo {author} {\bibfnamefont {I.~V.}\ \bibnamefont
  {Gornyi}}, \bibinfo {author} {\bibfnamefont {A.~D.}\ \bibnamefont {Mirlin}},
  \ and\ \bibinfo {author} {\bibfnamefont {D.~G.}\ \bibnamefont {Polyakov}},\
  }\href {\doibase 10.1103/PhysRevLett.95.206603} {\bibfield  {journal}
  {\bibinfo  {journal} {Phys. Rev. Lett.}\ }\textbf {\bibinfo {volume} {95}},\
  \bibinfo {pages} {206603} (\bibinfo {year} {2005})}\BibitemShut {NoStop}%
\bibitem [{\citenamefont {Nandkishore}\ and\ \citenamefont
  {Huse}(2015)}]{nandkishore2015many}%
  \BibitemOpen
  \bibfield  {author} {\bibinfo {author} {\bibfnamefont {R.}~\bibnamefont
  {Nandkishore}}\ and\ \bibinfo {author} {\bibfnamefont {D.~A.}\ \bibnamefont
  {Huse}},\ }\href@noop {} {\bibfield  {journal} {\bibinfo  {journal} {Annu.
  Rev. Condens. Matter Phys.}\ }\textbf {\bibinfo {volume} {6}},\ \bibinfo
  {pages} {15} (\bibinfo {year} {2015})}\BibitemShut {NoStop}%
\bibitem [{\citenamefont {Anderson}(1958)}]{PhysRev.109.1492}%
  \BibitemOpen
  \bibfield  {author} {\bibinfo {author} {\bibfnamefont {P.~W.}\ \bibnamefont
  {Anderson}},\ }\href {\doibase 10.1103/PhysRev.109.1492} {\bibfield
  {journal} {\bibinfo  {journal} {Phys. Rev.}\ }\textbf {\bibinfo {volume}
  {109}},\ \bibinfo {pages} {1492} (\bibinfo {year} {1958})}\BibitemShut
  {NoStop}%
\bibitem [{\citenamefont {Schulz}\ \emph {et~al.}(2019)\citenamefont {Schulz},
  \citenamefont {Hooley}, \citenamefont {Moessner},\ and\ \citenamefont
  {Pollmann}}]{schulz2019stark}%
  \BibitemOpen
  \bibfield  {author} {\bibinfo {author} {\bibfnamefont {M.}~\bibnamefont
  {Schulz}}, \bibinfo {author} {\bibfnamefont {C.}~\bibnamefont {Hooley}},
  \bibinfo {author} {\bibfnamefont {R.}~\bibnamefont {Moessner}}, \ and\
  \bibinfo {author} {\bibfnamefont {F.}~\bibnamefont {Pollmann}},\ }\href@noop
  {} {\bibfield  {journal} {\bibinfo  {journal} {Physical review letters}\
  }\textbf {\bibinfo {volume} {122}},\ \bibinfo {pages} {040606} (\bibinfo
  {year} {2019})}\BibitemShut {NoStop}%
\bibitem [{\citenamefont {van Nieuwenburg}\ \emph {et~al.}(2019)\citenamefont
  {van Nieuwenburg}, \citenamefont {Baum},\ and\ \citenamefont
  {Refael}}]{van2019bloch}%
  \BibitemOpen
  \bibfield  {author} {\bibinfo {author} {\bibfnamefont {E.}~\bibnamefont {van
  Nieuwenburg}}, \bibinfo {author} {\bibfnamefont {Y.}~\bibnamefont {Baum}}, \
  and\ \bibinfo {author} {\bibfnamefont {G.}~\bibnamefont {Refael}},\
  }\href@noop {} {\bibfield  {journal} {\bibinfo  {journal} {Proceedings of the
  National Academy of Sciences}\ ,\ \bibinfo {pages} {201819316}} (\bibinfo
  {year} {2019})}\BibitemShut {NoStop}%
\bibitem [{\citenamefont {Bhakuni}\ and\ \citenamefont
  {Sharma}(2019)}]{bhakuni2019entanglement}%
  \BibitemOpen
  \bibfield  {author} {\bibinfo {author} {\bibfnamefont {D.~S.}\ \bibnamefont
  {Bhakuni}}\ and\ \bibinfo {author} {\bibfnamefont {A.}~\bibnamefont
  {Sharma}},\ }\href@noop {} {\bibfield  {journal} {\bibinfo  {journal} {arXiv
  preprint arXiv:1909.10542}\ } (\bibinfo {year} {2019})}\BibitemShut {NoStop}%
\bibitem [{\citenamefont {Taylor}\ \emph {et~al.}(2019)\citenamefont {Taylor},
  \citenamefont {Schulz}, \citenamefont {Pollmann},\ and\ \citenamefont
  {Moessner}}]{taylor2019experimental}%
  \BibitemOpen
  \bibfield  {author} {\bibinfo {author} {\bibfnamefont {S.~R.}\ \bibnamefont
  {Taylor}}, \bibinfo {author} {\bibfnamefont {M.}~\bibnamefont {Schulz}},
  \bibinfo {author} {\bibfnamefont {F.}~\bibnamefont {Pollmann}}, \ and\
  \bibinfo {author} {\bibfnamefont {R.}~\bibnamefont {Moessner}},\ }\href@noop
  {} {\bibfield  {journal} {\bibinfo  {journal} {arXiv preprint
  arXiv:1910.01154}\ } (\bibinfo {year} {2019})}\BibitemShut {NoStop}%
\bibitem [{\citenamefont {Abanin}\ \emph {et~al.}(2015)\citenamefont {Abanin},
  \citenamefont {De~Roeck},\ and\ \citenamefont
  {Huveneers}}]{abanin2015exponentially}%
  \BibitemOpen
  \bibfield  {author} {\bibinfo {author} {\bibfnamefont {D.~A.}\ \bibnamefont
  {Abanin}}, \bibinfo {author} {\bibfnamefont {W.}~\bibnamefont {De~Roeck}}, \
  and\ \bibinfo {author} {\bibfnamefont {F.}~\bibnamefont {Huveneers}},\
  }\href@noop {} {\bibfield  {journal} {\bibinfo  {journal} {Physical review
  letters}\ }\textbf {\bibinfo {volume} {115}},\ \bibinfo {pages} {256803}
  (\bibinfo {year} {2015})}\BibitemShut {NoStop}%
\bibitem [{\citenamefont {D’Alessio}\ and\ \citenamefont
  {Rigol}(2014)}]{d2014long}%
  \BibitemOpen
  \bibfield  {author} {\bibinfo {author} {\bibfnamefont {L.}~\bibnamefont
  {D’Alessio}}\ and\ \bibinfo {author} {\bibfnamefont {M.}~\bibnamefont
  {Rigol}},\ }\href@noop {} {\bibfield  {journal} {\bibinfo  {journal}
  {Physical Review X}\ }\textbf {\bibinfo {volume} {4}},\ \bibinfo {pages}
  {041048} (\bibinfo {year} {2014})}\BibitemShut {NoStop}%
\bibitem [{\citenamefont {Lazarides}\ \emph
  {et~al.}(2014{\natexlab{a}})\citenamefont {Lazarides}, \citenamefont {Das},\
  and\ \citenamefont {Moessner}}]{lazarides2014equilibrium}%
  \BibitemOpen
  \bibfield  {author} {\bibinfo {author} {\bibfnamefont {A.}~\bibnamefont
  {Lazarides}}, \bibinfo {author} {\bibfnamefont {A.}~\bibnamefont {Das}}, \
  and\ \bibinfo {author} {\bibfnamefont {R.}~\bibnamefont {Moessner}},\
  }\href@noop {} {\bibfield  {journal} {\bibinfo  {journal} {Physical Review
  E}\ }\textbf {\bibinfo {volume} {90}},\ \bibinfo {pages} {012110} (\bibinfo
  {year} {2014}{\natexlab{a}})}\BibitemShut {NoStop}%
\bibitem [{\citenamefont {Lazarides}\ \emph
  {et~al.}(2014{\natexlab{b}})\citenamefont {Lazarides}, \citenamefont {Das},\
  and\ \citenamefont {Moessner}}]{lazarides2014periodic}%
  \BibitemOpen
  \bibfield  {author} {\bibinfo {author} {\bibfnamefont {A.}~\bibnamefont
  {Lazarides}}, \bibinfo {author} {\bibfnamefont {A.}~\bibnamefont {Das}}, \
  and\ \bibinfo {author} {\bibfnamefont {R.}~\bibnamefont {Moessner}},\
  }\href@noop {} {\bibfield  {journal} {\bibinfo  {journal} {Physical review
  letters}\ }\textbf {\bibinfo {volume} {112}},\ \bibinfo {pages} {150401}
  (\bibinfo {year} {2014}{\natexlab{b}})}\BibitemShut {NoStop}%
\bibitem [{\citenamefont {Ray}\ \emph {et~al.}(2018{\natexlab{a}})\citenamefont
  {Ray}, \citenamefont {Ghosh},\ and\ \citenamefont {Sinha}}]{ray2018drive}%
  \BibitemOpen
  \bibfield  {author} {\bibinfo {author} {\bibfnamefont {S.}~\bibnamefont
  {Ray}}, \bibinfo {author} {\bibfnamefont {A.}~\bibnamefont {Ghosh}}, \ and\
  \bibinfo {author} {\bibfnamefont {S.}~\bibnamefont {Sinha}},\ }\href@noop {}
  {\bibfield  {journal} {\bibinfo  {journal} {Physical Review E}\ }\textbf
  {\bibinfo {volume} {97}},\ \bibinfo {pages} {010101} (\bibinfo {year}
  {2018}{\natexlab{a}})}\BibitemShut {NoStop}%
\bibitem [{\citenamefont {Ray}\ \emph {et~al.}(2018{\natexlab{b}})\citenamefont
  {Ray}, \citenamefont {Sinha},\ and\ \citenamefont
  {Sengupta}}]{ray2018signature}%
  \BibitemOpen
  \bibfield  {author} {\bibinfo {author} {\bibfnamefont {S.}~\bibnamefont
  {Ray}}, \bibinfo {author} {\bibfnamefont {S.}~\bibnamefont {Sinha}}, \ and\
  \bibinfo {author} {\bibfnamefont {K.}~\bibnamefont {Sengupta}},\ }\href@noop
  {} {\bibfield  {journal} {\bibinfo  {journal} {Physical Review A}\ }\textbf
  {\bibinfo {volume} {98}},\ \bibinfo {pages} {053631} (\bibinfo {year}
  {2018}{\natexlab{b}})}\BibitemShut {NoStop}%
\bibitem [{\citenamefont {Abanin}\ \emph
  {et~al.}(2017{\natexlab{a}})\citenamefont {Abanin}, \citenamefont {De~Roeck},
  \citenamefont {Ho},\ and\ \citenamefont {Huveneers}}]{abanin2017effective}%
  \BibitemOpen
  \bibfield  {author} {\bibinfo {author} {\bibfnamefont {D.~A.}\ \bibnamefont
  {Abanin}}, \bibinfo {author} {\bibfnamefont {W.}~\bibnamefont {De~Roeck}},
  \bibinfo {author} {\bibfnamefont {W.~W.}\ \bibnamefont {Ho}}, \ and\ \bibinfo
  {author} {\bibfnamefont {F.}~\bibnamefont {Huveneers}},\ }\href@noop {}
  {\bibfield  {journal} {\bibinfo  {journal} {Physical Review B}\ }\textbf
  {\bibinfo {volume} {95}},\ \bibinfo {pages} {014112} (\bibinfo {year}
  {2017}{\natexlab{a}})}\BibitemShut {NoStop}%
\bibitem [{\citenamefont {Weidinger}\ and\ \citenamefont
  {Knap}(2017)}]{weidinger2017floquet}%
  \BibitemOpen
  \bibfield  {author} {\bibinfo {author} {\bibfnamefont {S.~A.}\ \bibnamefont
  {Weidinger}}\ and\ \bibinfo {author} {\bibfnamefont {M.}~\bibnamefont
  {Knap}},\ }\href@noop {} {\bibfield  {journal} {\bibinfo  {journal}
  {Scientific reports}\ }\textbf {\bibinfo {volume} {7}},\ \bibinfo {pages} {1}
  (\bibinfo {year} {2017})}\BibitemShut {NoStop}%
\bibitem [{\citenamefont {G\'omez-Le\'on}\ and\ \citenamefont
  {Platero}(2013)}]{PhysRevLett.110.200403}%
  \BibitemOpen
  \bibfield  {author} {\bibinfo {author} {\bibfnamefont {A.}~\bibnamefont
  {G\'omez-Le\'on}}\ and\ \bibinfo {author} {\bibfnamefont {G.}~\bibnamefont
  {Platero}},\ }\href {\doibase 10.1103/PhysRevLett.110.200403} {\bibfield
  {journal} {\bibinfo  {journal} {Phys. Rev. Lett.}\ }\textbf {\bibinfo
  {volume} {110}},\ \bibinfo {pages} {200403} (\bibinfo {year}
  {2013})}\BibitemShut {NoStop}%
\bibitem [{\citenamefont {Cayssol}\ \emph {et~al.}(2013)\citenamefont
  {Cayssol}, \citenamefont {D{\'o}ra}, \citenamefont {Simon},\ and\
  \citenamefont {Moessner}}]{cayssol2013floquet}%
  \BibitemOpen
  \bibfield  {author} {\bibinfo {author} {\bibfnamefont {J.}~\bibnamefont
  {Cayssol}}, \bibinfo {author} {\bibfnamefont {B.}~\bibnamefont {D{\'o}ra}},
  \bibinfo {author} {\bibfnamefont {F.}~\bibnamefont {Simon}}, \ and\ \bibinfo
  {author} {\bibfnamefont {R.}~\bibnamefont {Moessner}},\ }\href@noop {}
  {\bibfield  {journal} {\bibinfo  {journal} {physica status solidi
  (RRL)--Rapid Research Letters}\ }\textbf {\bibinfo {volume} {7}},\ \bibinfo
  {pages} {101} (\bibinfo {year} {2013})}\BibitemShut {NoStop}%
\bibitem [{\citenamefont {Rudner}\ \emph {et~al.}(2013)\citenamefont {Rudner},
  \citenamefont {Lindner}, \citenamefont {Berg},\ and\ \citenamefont
  {Levin}}]{rudner2013anomalous}%
  \BibitemOpen
  \bibfield  {author} {\bibinfo {author} {\bibfnamefont {M.~S.}\ \bibnamefont
  {Rudner}}, \bibinfo {author} {\bibfnamefont {N.~H.}\ \bibnamefont {Lindner}},
  \bibinfo {author} {\bibfnamefont {E.}~\bibnamefont {Berg}}, \ and\ \bibinfo
  {author} {\bibfnamefont {M.}~\bibnamefont {Levin}},\ }\href@noop {}
  {\bibfield  {journal} {\bibinfo  {journal} {Physical Review X}\ }\textbf
  {\bibinfo {volume} {3}},\ \bibinfo {pages} {031005} (\bibinfo {year}
  {2013})}\BibitemShut {NoStop}%
\bibitem [{\citenamefont {Else}\ \emph {et~al.}(2016)\citenamefont {Else},
  \citenamefont {Bauer},\ and\ \citenamefont {Nayak}}]{else2016floquet}%
  \BibitemOpen
  \bibfield  {author} {\bibinfo {author} {\bibfnamefont {D.~V.}\ \bibnamefont
  {Else}}, \bibinfo {author} {\bibfnamefont {B.}~\bibnamefont {Bauer}}, \ and\
  \bibinfo {author} {\bibfnamefont {C.}~\bibnamefont {Nayak}},\ }\href@noop {}
  {\bibfield  {journal} {\bibinfo  {journal} {Physical review letters}\
  }\textbf {\bibinfo {volume} {117}},\ \bibinfo {pages} {090402} (\bibinfo
  {year} {2016})}\BibitemShut {NoStop}%
\bibitem [{\citenamefont {Ponte}\ \emph
  {et~al.}(2015{\natexlab{a}})\citenamefont {Ponte}, \citenamefont {Chandran},
  \citenamefont {Papi{\'c}},\ and\ \citenamefont
  {Abanin}}]{ponte2015periodically}%
  \BibitemOpen
  \bibfield  {author} {\bibinfo {author} {\bibfnamefont {P.}~\bibnamefont
  {Ponte}}, \bibinfo {author} {\bibfnamefont {A.}~\bibnamefont {Chandran}},
  \bibinfo {author} {\bibfnamefont {Z.}~\bibnamefont {Papi{\'c}}}, \ and\
  \bibinfo {author} {\bibfnamefont {D.~A.}\ \bibnamefont {Abanin}},\
  }\href@noop {} {\bibfield  {journal} {\bibinfo  {journal} {Annals of
  Physics}\ }\textbf {\bibinfo {volume} {353}},\ \bibinfo {pages} {196}
  (\bibinfo {year} {2015}{\natexlab{a}})}\BibitemShut {NoStop}%
\bibitem [{\citenamefont {Lazarides}\ \emph {et~al.}(2015)\citenamefont
  {Lazarides}, \citenamefont {Das},\ and\ \citenamefont
  {Moessner}}]{lazarides2015fate}%
  \BibitemOpen
  \bibfield  {author} {\bibinfo {author} {\bibfnamefont {A.}~\bibnamefont
  {Lazarides}}, \bibinfo {author} {\bibfnamefont {A.}~\bibnamefont {Das}}, \
  and\ \bibinfo {author} {\bibfnamefont {R.}~\bibnamefont {Moessner}},\
  }\href@noop {} {\bibfield  {journal} {\bibinfo  {journal} {Physical review
  letters}\ }\textbf {\bibinfo {volume} {115}},\ \bibinfo {pages} {030402}
  (\bibinfo {year} {2015})}\BibitemShut {NoStop}%
\bibitem [{\citenamefont {Abanin}\ \emph {et~al.}(2016)\citenamefont {Abanin},
  \citenamefont {De~Roeck},\ and\ \citenamefont
  {Huveneers}}]{abanin2016theory}%
  \BibitemOpen
  \bibfield  {author} {\bibinfo {author} {\bibfnamefont {D.~A.}\ \bibnamefont
  {Abanin}}, \bibinfo {author} {\bibfnamefont {W.}~\bibnamefont {De~Roeck}}, \
  and\ \bibinfo {author} {\bibfnamefont {F.}~\bibnamefont {Huveneers}},\
  }\href@noop {} {\bibfield  {journal} {\bibinfo  {journal} {Annals of
  Physics}\ }\textbf {\bibinfo {volume} {372}},\ \bibinfo {pages} {1} (\bibinfo
  {year} {2016})}\BibitemShut {NoStop}%
\bibitem [{\citenamefont {D’Alessio}\ and\ \citenamefont
  {Polkovnikov}(2013)}]{d2013many}%
  \BibitemOpen
  \bibfield  {author} {\bibinfo {author} {\bibfnamefont {L.}~\bibnamefont
  {D’Alessio}}\ and\ \bibinfo {author} {\bibfnamefont {A.}~\bibnamefont
  {Polkovnikov}},\ }\href@noop {} {\bibfield  {journal} {\bibinfo  {journal}
  {Annals of Physics}\ }\textbf {\bibinfo {volume} {333}},\ \bibinfo {pages}
  {19} (\bibinfo {year} {2013})}\BibitemShut {NoStop}%
\bibitem [{\citenamefont {Ponte}\ \emph
  {et~al.}(2015{\natexlab{b}})\citenamefont {Ponte}, \citenamefont {Papi{\'c}},
  \citenamefont {Huveneers},\ and\ \citenamefont {Abanin}}]{ponte2015many}%
  \BibitemOpen
  \bibfield  {author} {\bibinfo {author} {\bibfnamefont {P.}~\bibnamefont
  {Ponte}}, \bibinfo {author} {\bibfnamefont {Z.}~\bibnamefont {Papi{\'c}}},
  \bibinfo {author} {\bibfnamefont {F.}~\bibnamefont {Huveneers}}, \ and\
  \bibinfo {author} {\bibfnamefont {D.~A.}\ \bibnamefont {Abanin}},\
  }\href@noop {} {\bibfield  {journal} {\bibinfo  {journal} {Physical review
  letters}\ }\textbf {\bibinfo {volume} {114}},\ \bibinfo {pages} {140401}
  (\bibinfo {year} {2015}{\natexlab{b}})}\BibitemShut {NoStop}%
\bibitem [{\citenamefont {Bordia}\ \emph {et~al.}(2017)\citenamefont {Bordia},
  \citenamefont {L{\"u}schen}, \citenamefont {Schneider}, \citenamefont
  {Knap},\ and\ \citenamefont {Bloch}}]{bordia2017periodically}%
  \BibitemOpen
  \bibfield  {author} {\bibinfo {author} {\bibfnamefont {P.}~\bibnamefont
  {Bordia}}, \bibinfo {author} {\bibfnamefont {H.}~\bibnamefont {L{\"u}schen}},
  \bibinfo {author} {\bibfnamefont {U.}~\bibnamefont {Schneider}}, \bibinfo
  {author} {\bibfnamefont {M.}~\bibnamefont {Knap}}, \ and\ \bibinfo {author}
  {\bibfnamefont {I.}~\bibnamefont {Bloch}},\ }\href@noop {} {\bibfield
  {journal} {\bibinfo  {journal} {Nature Physics}\ }\textbf {\bibinfo {volume}
  {13}},\ \bibinfo {pages} {460} (\bibinfo {year} {2017})}\BibitemShut
  {NoStop}%
\bibitem [{\citenamefont {Singh}\ \emph {et~al.}(2019)\citenamefont {Singh},
  \citenamefont {Fujiwara}, \citenamefont {Geiger}, \citenamefont {Simmons},
  \citenamefont {Lipatov}, \citenamefont {Cao}, \citenamefont {Dotti},
  \citenamefont {Rajagopal}, \citenamefont {Senaratne}, \citenamefont
  {Shimasaki} \emph {et~al.}}]{singh2019quantifying}%
  \BibitemOpen
  \bibfield  {author} {\bibinfo {author} {\bibfnamefont {K.}~\bibnamefont
  {Singh}}, \bibinfo {author} {\bibfnamefont {C.~J.}\ \bibnamefont {Fujiwara}},
  \bibinfo {author} {\bibfnamefont {Z.~A.}\ \bibnamefont {Geiger}}, \bibinfo
  {author} {\bibfnamefont {E.~Q.}\ \bibnamefont {Simmons}}, \bibinfo {author}
  {\bibfnamefont {M.}~\bibnamefont {Lipatov}}, \bibinfo {author} {\bibfnamefont
  {A.}~\bibnamefont {Cao}}, \bibinfo {author} {\bibfnamefont {P.}~\bibnamefont
  {Dotti}}, \bibinfo {author} {\bibfnamefont {S.~V.}\ \bibnamefont
  {Rajagopal}}, \bibinfo {author} {\bibfnamefont {R.}~\bibnamefont
  {Senaratne}}, \bibinfo {author} {\bibfnamefont {T.}~\bibnamefont
  {Shimasaki}},  \emph {et~al.},\ }\href@noop {} {\bibfield  {journal}
  {\bibinfo  {journal} {Physical Review X}\ }\textbf {\bibinfo {volume} {9}},\
  \bibinfo {pages} {041021} (\bibinfo {year} {2019})}\BibitemShut {NoStop}%
\bibitem [{\citenamefont {Rubio-Abadal}\ \emph {et~al.}(2020)\citenamefont
  {Rubio-Abadal}, \citenamefont {Ippoliti}, \citenamefont {Hollerith},
  \citenamefont {Wei}, \citenamefont {Rui}, \citenamefont {Sondhi},
  \citenamefont {Khemani}, \citenamefont {Gross},\ and\ \citenamefont
  {Bloch}}]{rubio2020floquet}%
  \BibitemOpen
  \bibfield  {author} {\bibinfo {author} {\bibfnamefont {A.}~\bibnamefont
  {Rubio-Abadal}}, \bibinfo {author} {\bibfnamefont {M.}~\bibnamefont
  {Ippoliti}}, \bibinfo {author} {\bibfnamefont {S.}~\bibnamefont {Hollerith}},
  \bibinfo {author} {\bibfnamefont {D.}~\bibnamefont {Wei}}, \bibinfo {author}
  {\bibfnamefont {J.}~\bibnamefont {Rui}}, \bibinfo {author} {\bibfnamefont
  {S.}~\bibnamefont {Sondhi}}, \bibinfo {author} {\bibfnamefont
  {V.}~\bibnamefont {Khemani}}, \bibinfo {author} {\bibfnamefont
  {C.}~\bibnamefont {Gross}}, \ and\ \bibinfo {author} {\bibfnamefont
  {I.}~\bibnamefont {Bloch}},\ }\href@noop {} {\bibfield  {journal} {\bibinfo
  {journal} {arXiv preprint arXiv:2001.08226}\ } (\bibinfo {year}
  {2020})}\BibitemShut {NoStop}%
\bibitem [{\citenamefont {Dunlap}\ and\ \citenamefont
  {Kenkre}(1986)}]{dunlap1986dynamic}%
  \BibitemOpen
  \bibfield  {author} {\bibinfo {author} {\bibfnamefont {D.}~\bibnamefont
  {Dunlap}}\ and\ \bibinfo {author} {\bibfnamefont {V.}~\bibnamefont
  {Kenkre}},\ }\href@noop {} {\bibfield  {journal} {\bibinfo  {journal}
  {Physical Review B}\ }\textbf {\bibinfo {volume} {34}},\ \bibinfo {pages}
  {3625} (\bibinfo {year} {1986})}\BibitemShut {NoStop}%
\bibitem [{\citenamefont {Dunlap}\ and\ \citenamefont
  {Kenkre}(1988)}]{dunlap1988dynamic}%
  \BibitemOpen
  \bibfield  {author} {\bibinfo {author} {\bibfnamefont {D.}~\bibnamefont
  {Dunlap}}\ and\ \bibinfo {author} {\bibfnamefont {V.}~\bibnamefont
  {Kenkre}},\ }\href@noop {} {\bibfield  {journal} {\bibinfo  {journal}
  {Physics Letters A}\ }\textbf {\bibinfo {volume} {127}},\ \bibinfo {pages}
  {438} (\bibinfo {year} {1988})}\BibitemShut {NoStop}%
\bibitem [{\citenamefont {Bhakuni}\ and\ \citenamefont
  {Sharma}(2018)}]{PhysRevB.98.045408}%
  \BibitemOpen
  \bibfield  {author} {\bibinfo {author} {\bibfnamefont {D.~S.}\ \bibnamefont
  {Bhakuni}}\ and\ \bibinfo {author} {\bibfnamefont {A.}~\bibnamefont
  {Sharma}},\ }\href {\doibase 10.1103/PhysRevB.98.045408} {\bibfield
  {journal} {\bibinfo  {journal} {Phys. Rev. B}\ }\textbf {\bibinfo {volume}
  {98}},\ \bibinfo {pages} {045408} (\bibinfo {year} {2018})}\BibitemShut
  {NoStop}%
\bibitem [{\citenamefont {Eckardt}\ \emph {et~al.}(2009)\citenamefont
  {Eckardt}, \citenamefont {Holthaus}, \citenamefont {Lignier}, \citenamefont
  {Zenesini}, \citenamefont {Ciampini}, \citenamefont {Morsch},\ and\
  \citenamefont {Arimondo}}]{eckardt2009exploring}%
  \BibitemOpen
  \bibfield  {author} {\bibinfo {author} {\bibfnamefont {A.}~\bibnamefont
  {Eckardt}}, \bibinfo {author} {\bibfnamefont {M.}~\bibnamefont {Holthaus}},
  \bibinfo {author} {\bibfnamefont {H.}~\bibnamefont {Lignier}}, \bibinfo
  {author} {\bibfnamefont {A.}~\bibnamefont {Zenesini}}, \bibinfo {author}
  {\bibfnamefont {D.}~\bibnamefont {Ciampini}}, \bibinfo {author}
  {\bibfnamefont {O.}~\bibnamefont {Morsch}}, \ and\ \bibinfo {author}
  {\bibfnamefont {E.}~\bibnamefont {Arimondo}},\ }\href@noop {} {\bibfield
  {journal} {\bibinfo  {journal} {Physical Review A}\ }\textbf {\bibinfo
  {volume} {79}},\ \bibinfo {pages} {013611} (\bibinfo {year}
  {2009})}\BibitemShut {NoStop}%
\bibitem [{\citenamefont {Bairey}\ \emph {et~al.}(2017)\citenamefont {Bairey},
  \citenamefont {Refael},\ and\ \citenamefont {Lindner}}]{bairey2017driving}%
  \BibitemOpen
  \bibfield  {author} {\bibinfo {author} {\bibfnamefont {E.}~\bibnamefont
  {Bairey}}, \bibinfo {author} {\bibfnamefont {G.}~\bibnamefont {Refael}}, \
  and\ \bibinfo {author} {\bibfnamefont {N.~H.}\ \bibnamefont {Lindner}},\
  }\href@noop {} {\bibfield  {journal} {\bibinfo  {journal} {Physical Review
  B}\ }\textbf {\bibinfo {volume} {96}},\ \bibinfo {pages} {020201} (\bibinfo
  {year} {2017})}\BibitemShut {NoStop}%
\bibitem [{\citenamefont {Holthaus}\ \emph
  {et~al.}(1995{\natexlab{a}})\citenamefont {Holthaus}, \citenamefont
  {Ristow},\ and\ \citenamefont {Hone}}]{holthaus1995random}%
  \BibitemOpen
  \bibfield  {author} {\bibinfo {author} {\bibfnamefont {M.}~\bibnamefont
  {Holthaus}}, \bibinfo {author} {\bibfnamefont {G.}~\bibnamefont {Ristow}}, \
  and\ \bibinfo {author} {\bibfnamefont {D.}~\bibnamefont {Hone}},\ }\href@noop
  {} {\bibfield  {journal} {\bibinfo  {journal} {EPL (Europhysics Letters)}\
  }\textbf {\bibinfo {volume} {32}},\ \bibinfo {pages} {241} (\bibinfo {year}
  {1995}{\natexlab{a}})}\BibitemShut {NoStop}%
\bibitem [{\citenamefont {Holthaus}\ \emph
  {et~al.}(1995{\natexlab{b}})\citenamefont {Holthaus}, \citenamefont
  {Ristow},\ and\ \citenamefont {Hone}}]{holthaus1995ac}%
  \BibitemOpen
  \bibfield  {author} {\bibinfo {author} {\bibfnamefont {M.}~\bibnamefont
  {Holthaus}}, \bibinfo {author} {\bibfnamefont {G.~H.}\ \bibnamefont
  {Ristow}}, \ and\ \bibinfo {author} {\bibfnamefont {D.~W.}\ \bibnamefont
  {Hone}},\ }\href@noop {} {\bibfield  {journal} {\bibinfo  {journal} {Physical
  review letters}\ }\textbf {\bibinfo {volume} {75}},\ \bibinfo {pages} {3914}
  (\bibinfo {year} {1995}{\natexlab{b}})}\BibitemShut {NoStop}%
\bibitem [{\citenamefont {Bhakuni}\ \emph {et~al.}(2019)\citenamefont
  {Bhakuni}, \citenamefont {Dattagupta},\ and\ \citenamefont
  {Sharma}}]{PhysRevB.99.155149}%
  \BibitemOpen
  \bibfield  {author} {\bibinfo {author} {\bibfnamefont {D.~S.}\ \bibnamefont
  {Bhakuni}}, \bibinfo {author} {\bibfnamefont {S.}~\bibnamefont {Dattagupta}},
  \ and\ \bibinfo {author} {\bibfnamefont {A.}~\bibnamefont {Sharma}},\ }\href
  {\doibase 10.1103/PhysRevB.99.155149} {\bibfield  {journal} {\bibinfo
  {journal} {Phys. Rev. B}\ }\textbf {\bibinfo {volume} {99}},\ \bibinfo
  {pages} {155149} (\bibinfo {year} {2019})}\BibitemShut {NoStop}%
\bibitem [{\citenamefont {Kudo}\ and\ \citenamefont
  {Monteiro}(2011)}]{kudo2011theoretical}%
  \BibitemOpen
  \bibfield  {author} {\bibinfo {author} {\bibfnamefont {K.}~\bibnamefont
  {Kudo}}\ and\ \bibinfo {author} {\bibfnamefont {T.}~\bibnamefont
  {Monteiro}},\ }\href@noop {} {\bibfield  {journal} {\bibinfo  {journal}
  {Physical Review A}\ }\textbf {\bibinfo {volume} {83}},\ \bibinfo {pages}
  {053627} (\bibinfo {year} {2011})}\BibitemShut {NoStop}%
\bibitem [{\citenamefont {Longhi}\ and\ \citenamefont
  {Della~Valle}(2012)}]{PhysRevB.86.075143}%
  \BibitemOpen
  \bibfield  {author} {\bibinfo {author} {\bibfnamefont {S.}~\bibnamefont
  {Longhi}}\ and\ \bibinfo {author} {\bibfnamefont {G.}~\bibnamefont
  {Della~Valle}},\ }\href {\doibase 10.1103/PhysRevB.86.075143} {\bibfield
  {journal} {\bibinfo  {journal} {Phys. Rev. B}\ }\textbf {\bibinfo {volume}
  {86}},\ \bibinfo {pages} {075143} (\bibinfo {year} {2012})}\BibitemShut
  {NoStop}%
\bibitem [{\citenamefont {Caetano}\ and\ \citenamefont
  {Lyra}(2011)}]{caetano2011wave}%
  \BibitemOpen
  \bibfield  {author} {\bibinfo {author} {\bibfnamefont {R.}~\bibnamefont
  {Caetano}}\ and\ \bibinfo {author} {\bibfnamefont {M.}~\bibnamefont {Lyra}},\
  }\href@noop {} {\bibfield  {journal} {\bibinfo  {journal} {Physics Letters
  A}\ }\textbf {\bibinfo {volume} {375}},\ \bibinfo {pages} {2770} (\bibinfo
  {year} {2011})}\BibitemShut {NoStop}%
\bibitem [{\citenamefont {Luitz}\ \emph {et~al.}(2017)\citenamefont {Luitz},
  \citenamefont {Bar~Lev},\ and\ \citenamefont {Lazarides}}]{luitz2017absence}%
  \BibitemOpen
  \bibfield  {author} {\bibinfo {author} {\bibfnamefont {D.~J.}\ \bibnamefont
  {Luitz}}, \bibinfo {author} {\bibfnamefont {Y.}~\bibnamefont {Bar~Lev}}, \
  and\ \bibinfo {author} {\bibfnamefont {A.}~\bibnamefont {Lazarides}},\
  }\href@noop {} {\bibfield  {journal} {\bibinfo  {journal} {SciPost Physics}\
  }\textbf {\bibinfo {volume} {3}},\ \bibinfo {pages} {029} (\bibinfo {year}
  {2017})}\BibitemShut {NoStop}%
\bibitem [{\citenamefont {Luitz}\ and\ \citenamefont
  {Lev}(2017)}]{luitz2017ergodic}%
  \BibitemOpen
  \bibfield  {author} {\bibinfo {author} {\bibfnamefont {D.~J.}\ \bibnamefont
  {Luitz}}\ and\ \bibinfo {author} {\bibfnamefont {Y.~B.}\ \bibnamefont
  {Lev}},\ }\href@noop {} {\bibfield  {journal} {\bibinfo  {journal} {Annalen
  der Physik}\ }\textbf {\bibinfo {volume} {529}},\ \bibinfo {pages} {1600350}
  (\bibinfo {year} {2017})}\BibitemShut {NoStop}%
\bibitem [{\citenamefont {{\v{Z}}nidari{\v{c}}}\ \emph
  {et~al.}(2008)\citenamefont {{\v{Z}}nidari{\v{c}}}, \citenamefont {Prosen},\
  and\ \citenamefont {Prelov{\v{s}}ek}}]{vznidarivc2008many}%
  \BibitemOpen
  \bibfield  {author} {\bibinfo {author} {\bibfnamefont {M.}~\bibnamefont
  {{\v{Z}}nidari{\v{c}}}}, \bibinfo {author} {\bibfnamefont {T.}~\bibnamefont
  {Prosen}}, \ and\ \bibinfo {author} {\bibfnamefont {P.}~\bibnamefont
  {Prelov{\v{s}}ek}},\ }\href@noop {} {\bibfield  {journal} {\bibinfo
  {journal} {Physical Review B}\ }\textbf {\bibinfo {volume} {77}},\ \bibinfo
  {pages} {064426} (\bibinfo {year} {2008})}\BibitemShut {NoStop}%
\bibitem [{\citenamefont {Bardarson}\ \emph {et~al.}(2012)\citenamefont
  {Bardarson}, \citenamefont {Pollmann},\ and\ \citenamefont
  {Moore}}]{PhysRevLett.109.017202}%
  \BibitemOpen
  \bibfield  {author} {\bibinfo {author} {\bibfnamefont {J.~H.}\ \bibnamefont
  {Bardarson}}, \bibinfo {author} {\bibfnamefont {F.}~\bibnamefont {Pollmann}},
  \ and\ \bibinfo {author} {\bibfnamefont {J.~E.}\ \bibnamefont {Moore}},\
  }\href {\doibase 10.1103/PhysRevLett.109.017202} {\bibfield  {journal}
  {\bibinfo  {journal} {Phys. Rev. Lett.}\ }\textbf {\bibinfo {volume} {109}},\
  \bibinfo {pages} {017202} (\bibinfo {year} {2012})}\BibitemShut {NoStop}%
\bibitem [{\citenamefont {Serbyn}\ \emph {et~al.}(2013)\citenamefont {Serbyn},
  \citenamefont {Papi\ifmmode~\acute{c}\else \'{c}\fi{}},\ and\ \citenamefont
  {Abanin}}]{PhysRevLett.110.260601}%
  \BibitemOpen
  \bibfield  {author} {\bibinfo {author} {\bibfnamefont {M.}~\bibnamefont
  {Serbyn}}, \bibinfo {author} {\bibfnamefont {Z.}~\bibnamefont
  {Papi\ifmmode~\acute{c}\else \'{c}\fi{}}}, \ and\ \bibinfo {author}
  {\bibfnamefont {D.~A.}\ \bibnamefont {Abanin}},\ }\href {\doibase
  10.1103/PhysRevLett.110.260601} {\bibfield  {journal} {\bibinfo  {journal}
  {Phys. Rev. Lett.}\ }\textbf {\bibinfo {volume} {110}},\ \bibinfo {pages}
  {260601} (\bibinfo {year} {2013})}\BibitemShut {NoStop}%
\bibitem [{\citenamefont {Mori}\ \emph {et~al.}(2016)\citenamefont {Mori},
  \citenamefont {Kuwahara},\ and\ \citenamefont {Saito}}]{mori2016rigorous}%
  \BibitemOpen
  \bibfield  {author} {\bibinfo {author} {\bibfnamefont {T.}~\bibnamefont
  {Mori}}, \bibinfo {author} {\bibfnamefont {T.}~\bibnamefont {Kuwahara}}, \
  and\ \bibinfo {author} {\bibfnamefont {K.}~\bibnamefont {Saito}},\
  }\href@noop {} {\bibfield  {journal} {\bibinfo  {journal} {Physical review
  letters}\ }\textbf {\bibinfo {volume} {116}},\ \bibinfo {pages} {120401}
  (\bibinfo {year} {2016})}\BibitemShut {NoStop}%
\bibitem [{\citenamefont {Abanin}\ \emph
  {et~al.}(2017{\natexlab{b}})\citenamefont {Abanin}, \citenamefont {De~Roeck},
  \citenamefont {Ho},\ and\ \citenamefont {Huveneers}}]{abanin2017rigorous}%
  \BibitemOpen
  \bibfield  {author} {\bibinfo {author} {\bibfnamefont {D.}~\bibnamefont
  {Abanin}}, \bibinfo {author} {\bibfnamefont {W.}~\bibnamefont {De~Roeck}},
  \bibinfo {author} {\bibfnamefont {W.~W.}\ \bibnamefont {Ho}}, \ and\ \bibinfo
  {author} {\bibfnamefont {F.}~\bibnamefont {Huveneers}},\ }\href@noop {}
  {\bibfield  {journal} {\bibinfo  {journal} {Communications in Mathematical
  Physics}\ }\textbf {\bibinfo {volume} {354}},\ \bibinfo {pages} {809}
  (\bibinfo {year} {2017}{\natexlab{b}})}\BibitemShut {NoStop}%
\bibitem [{\citenamefont {Luitz}\ \emph {et~al.}(2019)\citenamefont {Luitz},
  \citenamefont {Moessner}, \citenamefont {Sondhi},\ and\ \citenamefont
  {Khemani}}]{luitz2019prethermalization}%
  \BibitemOpen
  \bibfield  {author} {\bibinfo {author} {\bibfnamefont {D.~J.}\ \bibnamefont
  {Luitz}}, \bibinfo {author} {\bibfnamefont {R.}~\bibnamefont {Moessner}},
  \bibinfo {author} {\bibfnamefont {S.}~\bibnamefont {Sondhi}}, \ and\ \bibinfo
  {author} {\bibfnamefont {V.}~\bibnamefont {Khemani}},\ }\href@noop {}
  {\bibfield  {journal} {\bibinfo  {journal} {arXiv preprint arXiv:1908.10371}\
  } (\bibinfo {year} {2019})}\BibitemShut {NoStop}%
\end{thebibliography}%
\appendix
\onecolumngrid 

\section{Non- interacting Case: Semi-classical Description }\label{A1}
The form of the combined dc and ac field can be written as
\begin{equation}\label{s1}
\mathcal{F}(t) = 
\begin{cases}
F + A & \text{for }0\le t< T/2\\    
F - A   & \text{for }T/2\le t< T,\\    
\end{cases}
\end{equation}
where $A$ and $T$ are the amplitude and the time-period of the drive respectively and $F$ is the static field strength. 

Firstly, we will consider the case where the static field is absent ($F=0$). The quasi-momentum in the presence of this time dependent field changes as:$ 
q_{k}(t)=k+\frac{1}{\hbar}\int_{0}^{t} d\tau \mathcal{F}(\tau).$
For square wave driving (Eq.~\ref{s1}), the expression  for the quasi-momentum can be solved as~\cite{eckardt2009exploring}:
\begin{equation}
q_{k}(t) = 
\begin{cases}
k+A(t-T/4)/\hbar & \text{for }0\le t< T/2\\    
k+A(3T/4-t)/\hbar   & \text{for }T/2\le t< T.\\    
\end{cases}
\end{equation}
The change in the quasi-momentum leads to a change in the dispersion,
which is now time-dependent ($E(k)=-2J\cos(q_k(t))$). Due to the
absence of energy conservation, we focus on the one-cycle average of
the quasi-energy, which is given by: $
\epsilon(k)=\frac{-2J}{T}\int_{0}^{T} \cos[q_{k}(t)]\  dt.$
Substituting the expression for $q_{k}(t)$ and solving the integral, we get
\begin{equation}
\epsilon(k)=-2J \text{sinc}\left(\frac{\pi K}{2}\right)\cos(k),
\end{equation}
where, $\text{sinc}(z)=\sin(z)/z$ and $K=A/\omega$. The quasi-energy band
collapses at the zeros of the function $\text{sinc}(\pi K/2)$, which
occurs when $K = K_c = 2\nu$, $\nu$ being any integer (Fig.~\ref{qeig}(a)). This is the
condition for dynamic localization.

For a combined ac and dc field the quasi-momentum can be expressed as: $
q_{k}(t)=k+Ft+\int_{0}^{t} d\tau F_{ac}(\tau)$

Due to the dc part, the quasi-momentum is no longer a periodic function. However, for the resonance condition $F=n\omega$, the quasi-momentum becomes a periodic function. Solving for the one cycle average of quasi-energy, we get
\begin{eqnarray}
\epsilon(k)=\frac{-2J}{T}\int_{0}^{T} \cos[q_{k}(t)]\  dt
= \frac{-2J}{T}\int_{0}^{T/2} \cos\left(k+n\omega t+ A(t-T/4)\right) \ dt  + \int_{T/2}^{T} \cos\left(k+n\omega t+ A(3T/4-t)\right)\ dt.
\end{eqnarray}
The integral can be solved to yield:
\begin{equation}\label{A14}
\epsilon(k)=-2J_{\text{eff}}\cos(k+\frac{n\pi}{2}),\ \ \text{and} \quad J_{\text{eff}} = J \left\lbrace\frac{\sin(\frac{n\pi}{2}+\frac{K\pi}{2})}{(K\pi+n\pi)}+(-1)^{n}\frac{\sin(-\frac{n\pi}{2}+\frac{K\pi}{2})}{(K\pi-n\pi)}\right\rbrace
\end{equation}
\subsection{Case 1: Odd $n$}
For odd $n$, Eqn.~\ref{A14} can be simplified to
\begin{equation}
\epsilon(k)=-2J\left(\frac{2K\cos(\frac{K\pi}{2})}{(K^{2}-n^2)\pi}\right)\cos(k-\frac{\pi}{2}).
\end{equation}
Here, the band collapses for $K=K_c=2\nu +1, \ \nu\in\mathbb{Z}$ and $K_{c}\neq n$ (Fig.~\ref{qeig}(b)). This gives the condition of dynamic localization whereas for other $K$ with the resonance condition destruction of Wannier-Stark localization occurs. 

\subsection{Case 2: Even $n$}
For even $n$, Eqn.~\ref{A14} can be simplified to
\begin{equation}
\epsilon(k)=-2J\left(\frac{2K\sin(\frac{K\pi}{2})}{(K^{2}-n^2)\pi}\right)\cos(k) =-2J\left(\frac{\text{sinc}(\frac{K\pi}{2})}{1-\frac{n^2}{K^{2}}}\right)\cos(k).
\end{equation}
Again, the band collapse occurs at $K=K_{c}=2\nu,\  \nu\in\mathbb{Z}$ and  $K_{c}\neq n$ (Fig.~\ref{qeig}(c)). At these points an initially localized wave packet returns to its starting position. This gives the condition of dynamical localization.
For other values of $K$, and provided that the resonance condition holds, band formation takes place and the WS localization due to the static dc field is destroyed.
\subsection{Super-Bloch Oscillations}
Considering the case of a slight detuning from the resonant condition: $F=(n+\delta)\omega$,
the corresponding quasi-momentum can be written as:
$q_{k}(t)=k+n\omega t+\delta\omega t+\int_{0}^{t} d\tau F_{ac}(\tau)$.
The quasi-momentum is no longer periodic due to the extra
term. However for $\delta <<1$, we can approximately take $q_{k}(t)$
as periodic and can proceed further to calculate the quasi-energy by
assuming $\delta\omega t$ as a constant. It can be easily verified that
for both even and odd $n$, the cosine term $\cos(k)$, acquires
an additional phase $\delta\omega t$, which is equivalent to a static
dc field of magnitude $\delta\omega $. The dynamics shows
oscillatory behaviour similar to Bloch oscillations. These
oscillations are termed as super-Bloch oscillations. The time period
is given by: $T_{\text{SBO}}=\frac{2\pi}{\delta\omega}$.

	\section{Undriven system: \boldmath $\mathcal{F}(t)=F$}\label{A2}
	In the absence of driving, the Hamiltonian can be written as 
	\begin{eqnarray}
	H=-J\sum_{j=0}^{L-2}(c_{j}^{\dagger}c_{j+1}+c_{j+1}^{\dagger}c_{j})-F\sum_{j=0}^{L-1} j (n_{j}-\frac{1}{2}) + \alpha \sum_{j=0}^{L-1} \frac{j^2}{(L-1)^2} (n_{j}-\frac{1}{2}) + V\sum_{j=0}^{L-2} (n_j-\frac{1}{2})(n_{j+1}-\frac{1}{2}).
	\end{eqnarray}
	The model shows an ergodic phase for small field strengths, while for sufficiently strong field strength it shows an MBL phase. 
	
\begin{figure}[!h]
	\includegraphics[scale=1.03]{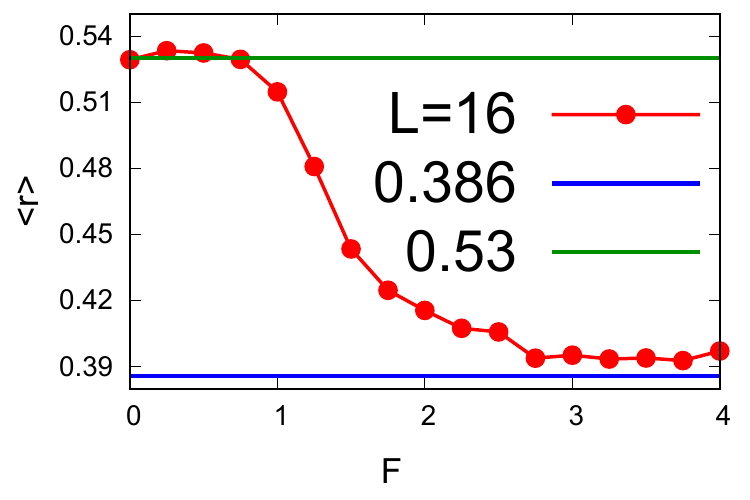}
	\includegraphics[scale=1.2]{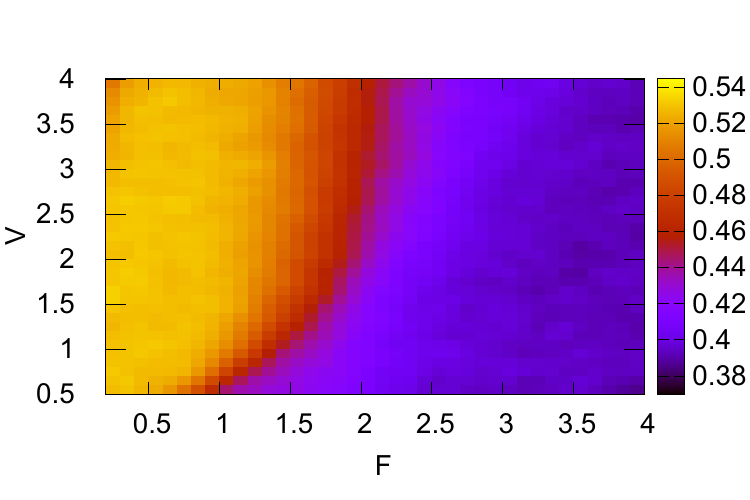}
	\caption{Left: Average level-spacing ratio as a function of the field strength for a fixed $V=1.0$. Right: Surface plot of the level statistics as a function of both field strength $(F)$ and the interaction strength $(V)$.  The other parameters are: $L=16, \alpha=1.0$. }
	\label{rav_sup}
\end{figure}

	To study the different phases, we plot the average level-spacing ratio
	as a function of field strength (Fig.~\ref{rav_sup}(a)). For small
	field strength the level-spacing follows Wigner-Dyson statistics while
	for large field strength it follows Poisson statistics signifying an
	MBL phase for these values of the field strength.  Finally, we also
	look at the interplay of the interactions and the field
	strength. Fig.~\ref{rav_sup}(b) shows the surface plot of the average
	level-spacing ratio as a function of both field strength and the
	interaction strength. On increasing the interaction strength, the
	field required to show MBL also increases.
\end{document}